\documentclass[10pt]{article}
\usepackage[a4paper, margin=1in]{geometry}

\usepackage[utf8]{inputenc}
\usepackage[]{graphicx}

\graphicspath{{figures/}}
\usepackage[export]{adjustbox}
\usepackage{graphbox}
\usepackage{caption}
\usepackage[final]{changes}
\definechangesauthor[color=orange]{ALB}
\definechangesauthor[color=blue]{OUS}
\definechangesauthor[color=red]{review}

\newcommand{\albdelete}[1]{}

\usepackage{todonotes}
\usepackage{amsmath}
\usepackage{amsfonts}
\usepackage{amssymb}
\usepackage{array}
\usepackage{longtable}
\usepackage{hyperref}
\usepackage[final]{showlabels}
\usepackage[percent]{overpic}

\usepackage{color}
\usepackage{enumitem}
\usepackage{authblk}

\title{{Navigating Local Minima and Bifurcations in Brittle Thin Film Systems with Irreversible Damage}}

\author[1]{M. M. Terzi}
\author[1,2]{O. U. Salman}
\author[1]{D. Faurie}
\author[3]{A. A. León Baldelli}
\affil[1]{LSPM, CNRS UPR3407, Universit\'e Sorbonne Paris Nord, 93400, Villateneuse, France}
\affil[2]{Lund University, Department of Mechanical Engineering Sciences, Lund, Sweden}
\affil[3]{Sorbonne Université, CNRS, Institut Jean Le Rond d'Alembert, F-75005 Paris, France}

\date{\today}

\begin{document}

\maketitle

\begin{abstract}
    \noindent
    {We investigate the computation of stable fracture paths in brittle thin films using one-dimensional damage models with an elastic foundation. The underlying variational formulation is non-convex, making the evolution path sensitive to algorithmic choices.}
    In this paper, we inquire into the effectiveness of quasi-Newton algorithms as an alternative to conventional Newton-Raphson solvers. These algorithms improve convergence by constructing a positive definite approximation of the Hessian, trading improved convergence with the risk of missing bifurcation points and stability thresholds.
    In the absence of irreversibility constraints, we construct an equilibrium map that represents all stable and unstable equilibrium states as a function of the external load, using well-known branch-following bifurcation techniques.
    Our main finding is that quasi-Newton algorithms fail to select stable evolution paths without exact  second variation information. {To overcome this, we introduce a spectral stability criterion based on the full Hessian, which enables the identification of optimal perturbations and improves path-following accuracy.}
        {We then extend our analysis to the irreversible case, where admissible perturbations are constrained to a cone. We develop a nonlinear constrained eigenvalue solver to compute the minimal eigenmode within this restricted space and show that it plays a key role in distinguishing physical instabilities from numerical artifacts. Our results provide practical guidance for robust computation of fracture paths in irreversible, non-convex settings.}
\end{abstract}

\newcommand{\Estiff}{\Psi}
\newcommand{\Ecompl}{\widetilde\Psi}
\newcommand{\subsu}{v}
\newcommand{\utest}{w}
\newcommand{\hfilm}{\mathsf h}
\newcommand{\hinter}{\mathsf{\hat{h}}}
\newcommand{\subsutest}{\tilde v}
\newcommand{\homogdiss}{\mathsf{w}}
\newcommand{\soften}{\mathsf{a}}
\newcommand{\damagell}{\ell}
\newcommand{\elastell}{\Lambda}
\newcommand{\stiffratio}{\rho}
\newcommand{\yhom}{y^{\text{hom}}}
\newcommand{\stiffmat}{\mathbf{H}}
\newcommand{\stiffmatcompl}{\widetilde{\mathbf{H}}}
\newcommand{\Youngfilm}{\mathsf{E_{\text{2d}}}}
\newcommand{\Youngsubs}{{\mathsf{\hat E_{\text{2d}}}}}
\newcommand{\Estiffhom}{\Estiff^\text{hom}}
\newcommand{\Ecomplhom}{\Ecompl^\text{hom}}
\newcommand{\femstate}{\mathbf{X}}
\newcommand{\femnewstate}{{\femstate^*}}
\newcommand{\femcurrentstate}{{\femstate_t}}
\newcommand{\femperturb}{\mathbf{p}}
\newcommand{\conespace}{K^+_{0}}

\begin{table}[h!]
    \centering
    \begin{tabular}{  m{6.cm}  m{3cm}  m{5.5cm}  }
        \hline
        \textbf{Description}                & \textbf{Symbol}                                    & \textbf{Remarks}                                                                \\
        \hline
        Energy stiff                        & $\Estiff$                                          &                                                                                 \\
        Energy compliant                    & $\Ecompl$                                          &                                                                                 \\
        Load                                & $\bar\epsilon_t$                                   &                                                                                 \\
        Load parameter                      & $t>0$                                              &                                                                                 \\
        Homogeneous energy, stiff           & $\Estiffhom$                                       &                                                                                 \\
        Homogeneous energy, compliant       & $\Ecomplhom$                                       &                                                                                 \\
        Material functions                  & $\mathsf E(\alpha), \homogdiss(\alpha)$            &                                                                                 \\
        Internal damage length              & $\bar\damagell$, $\ell$                            & dimensional, [m] (resp. nondimensional)                                         \\
        Film thickness                      & $h$                                                & dimensional, [m]                                                                \\
        Characteristic size                 & $L$                                                & dimensional, [m]                                                                \\
        Young modulus, substrate            & $\Youngsubs$                                       &                                                                                 \\
        Young modulus, film                 & $\Youngfilm$                                       &                                                                                 \\
        State, stiff                        & $(u, \alpha)$                                      &                                                                                 \\
        State, compliant                    & $(u, \alpha, v)$                                   &                                                                                 \\
        Equilibrium state, stiff            & $y_t=(u_t, \alpha_t)$                              &                                                                                 \\
        Equilibrium state, compliant        & $y_t=(u_t, \alpha_t, v_t)$                         &                                                                                 \\
        Homogeneous equilibrium state       & $\yhom$                                            &                                                                                 \\
        State, perturbations (stiff)        & $(u,\alpha), (\utest, \beta)$                      &                                                                                 \\
        State, perturbations (compliant)    & $(u, \alpha, \subsu), (\utest, \beta, \subsutest)$ &                                                                                 \\
        Admissible perturbations, stiff     & $V$                                                & $H^1(0,1)\times  H^1(0,1)$                                                      \\
        Admissible perturbations, compliant & $\widetilde{V}$                                    & $H^1(0,1)\times  H^1(0,1)\times  H^1(0,1)$                                      \\
        State space, stiff                  & $X_t$                                              & $H^1_t(0,1)\times  H^1(0,1)$                                                    \\
        State space, compliant              & $\widetilde{X_t}$                                  & $H^1_t(0,1)\times  H^1(0,1)$                                                    \\
        Homogeneous space                   & $V_0$                                              & $H^1_0(0,1)\times  H^1(0,1)$                                                    \\
        Cone of admissible perturbations    & $\conespace$                                       & $H^1_0(0, 1)\times \{w \in H^1((0, 1)): w(x) \geq 0 \text{ a.e. }x\in (0, 1)\}$ \\
        Dual Cone                           & $K^*$                                              & $\{y \in H^1(0,1): \langle x,y \rangle \leq 0, \forall x \in \conespace\}$      \\
        Eigenvalue, eigenfunction           & $(\lambda, w^*)$                                   &                                                                                 \\
        First critical load                 & $\bar \epsilon^c_1$                                &                                                                                 \\
        Critical loads                      & $\bar \epsilon^c_*$                                &                                                                                 \\
        $n$-th critical load                & $\bar \epsilon^c_n$                                &                                                                                 \\
        FEM Residual vectors                & ${\bf R}, \widetilde{\bf R}$                       &                                                                                 \\
        FEM base functions and derivatives  & ${\mathcal N}_i, {\mathcal N}'_i$                  &                                                                                 \\
        Eigenvalue at load $t$              & $\lambda_t$                                        &                                                                                 \\
        FEM state, generic                  & $\femstate$                                        &                                                                                 \\
        FEM new state                       & $\femnewstate$                                     &                                                                                 \\
        FEM current equilibrium state       & $\femcurrentstate$                                 &                                                                                 \\
        FEM perturbation                    & $\femperturb$                                      &                                                                                 \\
    \end{tabular}
    \caption{Notation and Definitions. The subscripted $t$ in $H_t^1(0, 1)$ indicates a $t$-parametrised boundary datum}
    \label{table:notation}
\end{table}

\section{Introduction}
 {
  Many physical phenomena in materials science such as crystal plasticity, phase transitions, twinning~\cite{Clayton2011-xq}, and fracture~\cite{francfort_marigo1998,Baldelli2014-ho,Baldelli2021-gc} can be modeled using nonlinear energy functionals at the mesoscale. These functionals describe the system's energy landscape in terms of configurational variables evolving under external loading. The evolution proceeds quasistatically through equilibrium states, typically obtained via incremental energy minimization along a prescribed loading path.
  Two broad classes of energy functionals are widely employed. The first, of the form $\Psi(u)$, depends solely on a kinematic (displacement) field $u$, and appears in theories such as quasi-continuum models, multi-well Landau-type theories of phase transformation, and crystal plasticity~\cite{Tadmor1996-qi,Lookman2003-gd,Conti2004-yj,Finel2010-zw,Salman2019-cg,Baggio2019-rs,Baggio2023-qu}. A second class, $\Psi(u,\alpha)$, introduces an internal scalar variable $\alpha$, often used in phase-field models to encode crystal structure, symmetry, orientation~\cite{Finel2010-zw,Ruffini2015-pn,Javanbakht2016-dr}, or damage~\cite{francfort_marigo1998,Salman2021-mn}.
  In both settings, the evolution problem involves searching for configurations that minimize the energy or satisfy optimality condition, potentially under constraints: $\min_{u} \Psi(u)$, or $\min_{u, \alpha} \Psi(u, \alpha)$.
  The nonconvex nature of these functionals typically leads to multiple local minima and competing equilibrium paths. Under quasi-static loading, the system transitions among metastable branches of equilibria, which may bifurcate, intersect, or terminate at loss-of-stability points. Near such instabilities, the system may switch abruptly to a new configuration, often involving dissipation.
  Understanding these transitions requires a careful analysis of bifurcation and stability, particularly to identify which equilibrium paths are physically relevant and observable. When stationarity alone does not suffice to determine the evolution due to multiple competing minima second-order conditions become essential.
  While classical bifurcation theory offers a well-developed framework for unconstrained dynamical systems—through canonical forms and linearized analysis near critical points~\cite{Iooss2012-el}, the situation is more intricate in the presence of internal variables and nonconvex constraints. In quasi-static evolutions governed by variational inequalities, as in damage or phase-field fracture models, bifurcation and stability are not solely determined by the existence of nearby equilibrium solutions in parameter space~\cite{Hill1958-xd,Bazant2010-zb}, because accessibility of states (restrained by constrained internal variables) plays a key role.
 }

\albdelete{In our context, a bifurcation condition along the system's evolution parametrized by the control parameter(s) is associated with  the uniqueness of a field of vectors tangent to the trajectory in phase space.}
{The fracture of brittle thin films bonded to substrates unveils a myriad of complex crack patterns, as evidenced by numerical studies \cite{Baldelli2014-ho,Alessi2019-bx,Hu2020-nt,Salman2021-mn,Baldelli2021-gc}, resembling those observed in natural contexts such as sand or dried mud \cite{Goehring2010-xz}, and even in biological structures like animal skin \cite{Qin2014-wz} and bark~\cite{chattaway:1955, shen:2020}. These phenomena hold particular relevance in the domain of stretchable and flexible electronics \cite{Faurie2019-to,Godard2022-ss} including  self-healing metal thin films on a flexible substrates \cite{Trost2024-ca}.}
In this work, we aim at characterising the stability (or observability) of static solutions (at a given control parameter) as well as to describe the evolutionary paths stemming from an initial condition. Conditions for uniqueness of the evolution path (or its non-bifurcation) reduce to the uniqueness of solutions to a boundary value problem defined for the \emph{rates of evolution}, or equivalently, the positive definiteness of its bilinear operator in a vector space.

    {
        The bifurcation of equilibrium solutions in unconstrained systems directly implies a change in stability of the solution which, physically, can lead to significant changes in material behavior or in the overall structural response.
    } Illustrative in this sense are Euler buckling~\cite{Bettiol2020-ey}, wrinkling in thin films~\cite{Hutchinson2013-jk}, homogeneous nucleation of dislocations in a crystal~\cite{Carpio2005-bv,Plans2007-cx,Baggio2019-rs,Mayer2022-km,Baggio2023-qu}, buckling of lattice structures~\cite{Combescure2016-dy,Bertoldi2008-au}, nucleation of cracks in soft solids or in pantographic structures \cite{Riccobelli2023-fc,Salman2021-mn}, plastic  avalanches in crystals or amorphous materials \cite{Zhang2020-ax,Weiss2021-db,Yang2020-zm}.
{When constraints are present, however, a stability transition is no longer implied by the loss of uniqueness of the solution, as the concept of stability must explicitly account for restrictions in the space of admissible variations.}
The absence of analytical solutions in strongly non-linear settings requires resorting to numerical methods for computing and predicting equilibrium configurations that correspond to the minima of an energy functional. Numerically, the minimization process involves discretizing the continuum fields onto a computational grid using methods such as finite elements, finite differences, or spectral techniques~\cite{Salman2009-qv,Liu2013-sj,Hu2021-mq}. Afterwards, an iterative solver is employed to seek equilibrium energy states, with options including the Newton-Raphson method~\cite{Wick2017-bo}, fixed-point iteration~\cite{Chen2019-mn,Kirkesaether_Brun2020-wa,Storvik2021-cd}, line-search-based descent algorithms like steepest descent or conjugate gradient~\cite{Stiefel1952-fw,Dai1999-hz}, quasi-Newton methods such as the highly-efficient Limited-memory Broyden-Fletcher-Goldfarb-Shanno (\textsc{L-BFGS}) approach~\cite{Liu1989-kl} which involves approximating the Hessian matrix, or more recent advancements like the fast inertial relaxation engine (\textsc{FIRE})~\cite{Guenole2020-tc}.
{
Recently, several authors~\cite{bharali:2022, najmeddine:2024, wu:2020} demonstrated the efficiency of the quasi-Newton BFGS
method for the fully coupled phase-field fracture problem. In particular, \cite{najmeddine:2024} reported reduction of computation
times by several orders of magnitude with the BFGS method, and the number of load iterations required by a staggered solution scheme as 3000 times higher than the number of iterations required by the BFGS method.
}
These solvers iteratively refine solutions starting from an initial guess provided as part of the solution procedure. Despite their widespread application, there remains a lack of clear understanding regarding the performance of these algorithms and their effectiveness in locating local minima.

    {We build upon prior work~\cite{Baldelli2021-gc}
        to investigate
        the numerical computation of stable solutions of phase-field fracture evolutionary models, extending the analysis to the full nonlinear stability of the solutions.
        Our investigation starts from the unconstrained case where damage and cracks can heal (without energy expense), which is unrealistic in the applications but allows to thorougly investigate the system's bifurcations and to highlight the main mathematical differences between the reversible and irreversible cases.
    }

    {We construct an \emph{equilibrium map} in this setup, allowing all  {inhomogeneous} solutions connected to the homogeneous branch to be identified along with their stability.
        This enables us to monitor the solutions returned by various numerical optimization techniques and assess their observability.
        The reversible case serves as a prototypical study which
        may still be relevant in certain phase-field damage models where irreversibility is imposed only on crack sets that exceed a \emph{given} damage threshold, referred to as `relaxed crack-set irreversibility' \cite{Bourdin2000-pc, Kumar2020-xz, De-Lorenzis2020-rz}, or  in models with softening elastic energy without irreversibility constraint \cite{Truskinovsky2010-st,Salman2019-kp,Salman2021-mn,Baggio2023-yo}.}

    {
        Our primary objective here is to study the interaction between numerical methods and variational structure in irreversibly-constrained systems using a physically motivated example that is both analytically tractable and rich in bifurcation behaviour. The brittle thin film on an elastic foundation provides a minimal yet non-trivial benchmark that captures the essential phenomenology: homogeneous to localised transitions, constraint activation, and numerical sensitivity, without requiring additional complexity from higher dimensions.
        Its `tractability' stems from its dimensional simplicity (our setup allows for a spatially constant stress), while its complexity arises from the elastic foundation, which introduces a lower-order perturbation in the energy landscape and enables a rich bifurcation structure.
    }
To this end,
we consider two one-dimensional phase-field fracture models of a brittle membrane on two types of substrates.\albdelete{: one stiff, one compliant.} The first model describes a brittle thin film deposited on a stiff substrate, while the second model involves a compliant yet unbreakable substrate that can undergo non-uniform deformations. The finite stiffness of the substrate in the second scenario leads to nontrivial qualitative differences in terms of uniqueness of the evolution path, associated with  the loss of stability of the unfractured solution~\cite{Baldelli2014-ho,Kuhn2015-rt,Baldelli2021-gc,Harandi2023-cd,zolesi:2024}.

Despite the one-dimensional setting we adopt here which allows for analytical predictions, these models reveal a complex landscape of equilibrium states with multiple local minima.
In the absence of an irreversibility constraint, bifurcation points from homogeneous solution can easily be calculated analytically and numerically, by employing continuation techniques.

Our findings indicate that under quasi-static loading conditions, line-search-based descent algorithms not relying on full Hessian can fail to detect expected branch-switching events and may return solutions that persist on unstable branches, thus lacking physical relevance.
We propose a remedy to this situation which involves utilizing information from the Hessian of the functional when it becomes singular.
To discuss this scenario we distinguish two settings, namely i) that in which damage is reversible and all small perturbations are admissible, and ii) the case where damage is subject to an irreversibility constraint which forbids healing. In the former scenario  negative variations of damage are allowed and indeed may occur - if convenient from an energetic viewpoint. In the second setting, instead, we consider damage as a unilateral irreversible process stemming from an irreducible one-directional pointwise growth constraint.

The rest of the paper is organized as follows. In Section \ref{sec:rigid}, we present one-dimensional phase-field fracture models with both rigid and compliant elastic foundations. In Section \ref{sec:stability} we focus on the analysis  of linear and non-linear stability {of homogeneous} solutions. In Section \ref{sec:numerics}, we construct the equilibrium map in the reversible setup, discuss the selection of equilibrium branches using various numerical optimization algorithms, and explore how irreversibility affects the stability of solutions. In the final Section \ref{sec:discussion}  we summarize our results.

\paragraph{Notation.} We employ standard notation for scalar Sobolev spaces defined on the unit interval, such as $H^1(0, 1)$, derivatives of one-dimensional fields, and matrix indices. We indicate the $L^2(0, 1)$-inner product of functions $u, v$ by $\langle u, v\rangle=\int_0^1 uv dx$. Subscripted $t$ means $t$-parametrised quantities, superscripted $(k)$ means $k$-th iterate of an iterative algorithm.
We indicate with boldface letters finite element matrices and vectors.
We use the prime sign to indicate spatial derivatives {and derivatives of functions of one variable, with respect to their argument}.
We use American English spelling throughout the text.
\section{Material, Structure, and Evolution}
\label{sec:rigid}
{We analyze the emergence of multiple equilibria, bifurcation phenomena, and stability transitions in two one-dimensional phase-field models of brittle thin films adhering to a substrate with distinct elastic behaviour. This modeling framework enables a systematic exploration of the interplay between loading conditions and the evolution of fracture patterns.}
\paragraph{Material Model}
We consider an isotropic and homogeneous brittle material {undergoing  in-plane displacements $u(x)$ and} modelled by an energy density (a state function) $W(e, \alpha, \alpha')$ which, at any point $x$,
depends on the local membrane strain $e(x)= u'(x)$, the local damage $\alpha(x)$, and its local gradient $\alpha'(x)$.
Here, the damage variable $\alpha$ is a scalar field driving material softening, bounded between 0 and 1, where $0$ indicates the undamaged material and $1$, the cracked material. Thus, at points where $\alpha=0$ the material is elastic with a stiffness $\mathsf{E}$ (its Young modulus), at points where $\alpha=1$ the material has a crack and zero residual stiffness, whereas for intermediate damage values the material's stiffness is $0<\mathsf{E}\soften(\alpha)< \mathsf{E}\mathsf{a}(0)$.
The state function $W$ is defined as
\begin{equation}
    \label{def:energy_material}
    W(e, \alpha, \alpha'):= \frac{1}{2} \mathsf{E} \soften(\alpha)e^2 +\mathsf{w_1}\homogdiss(\alpha) + {\mathsf{w_1}}\frac{\damagell^2}{2}\alpha'^2,
\end{equation}
where in the first summand, $\soften(\alpha)$ is the function that describes the material softening. On the other hand, $\homogdiss(\alpha)$ can be interpreted as the (normalised) energy dissipated during an homogeneous damaging process. It is combined with a term proportional to the square of its gradient which controls the energy cost of spatial damage variations.
    {The coefficient $\mathsf{w_1}$ has the dimensions of an energy density (J/m$^3$) and represents the material's damage dissipation rate, the product $\mathsf{w_1} \ell$ is dimensionally equivalent to a fracture toughness $G_c$, justifying the scaling $\mathsf{w_1} \sim G_c / \ell$.
        For a uniaxial damage localization zone of characteristic length $\ell$, this identification provides consistency between the continuum model and the energy released per unit crack surface in classical fracture mechanics.}
For physical consistency, $\soften(\alpha)$ is a non-negative function monotonically decreasing from 1 as $\alpha$ increases, reaching zero for $\alpha=1$.
On the other hand, $\homogdiss(\alpha)$ is non-negative, zero only if $\alpha=0$, and monotonically increasing with $\alpha$, reaching $\homogdiss(1)=1$.
The damage-dependent stress is $\sigma(\alpha):=\mathsf{E}a(\alpha) e$. The parameter $\damagell$ is a characteristic length that controls the competition between localisation and homogeneous damage, effectively controlling the width of damage localisations, the peak stress of the material in one-dimensional traction experiments, and - more in general - structural size effects.
Specifically, both functions $\soften(\alpha)$
and $\homogdiss(\alpha)$ are chosen to be quadratic, namely
\begin{equation}
    \label{def:constitutive_functions}
    \soften(\alpha) = (1-\alpha)^2, \quad \homogdiss(\alpha) = \alpha^2,
\end{equation}
This modelling choice is common (yet not unique) in phase-field fracture models (cf.~\cite{Bourdin2000-pc,Miehe2010-sj,Miehe2010-ja}). In the current context, it allows damage to evolve for an arbitrarily small value of the load. {
        The choice of a quadratic damage potential introduces the fundamental conceptual (and computational) difficulty of the evolution law already at the initial loading step.
        In contrast, the commonly used model with $w(\alpha)$ linear in $\alpha$ (usually referred to as AT1)  features an elastic regime in which damage variations are ruled out by optimality.
        Consequently, second order conditions (bifurcation and stability) are trivial, because damage cannot evolve.
        In our quadratic model, instead, damage can evolve from the very first discrete load increment, making the space of admissible damage variations nontrivial at all positive load levels. Specifically, in the irreversible scenario, admissible perturbations constitute a {cone with nonempty interior}, requiring careful handling of the inherent nonlinearity at both first and second orders. Thus, our model choice highlights a significant numerical challenge arising immediately at the initial time step, clearly distinguishing reversible from irreversible stability and bifurcation problems.
        Furthermore, this model naturally enforces the physical bounds on the damage variable (0 and 1) through energy minimality alone. Hence, no additional numerical constraints are required at first order, greatly simplifying numerical implementations and enabling exploration of the complex bifurcation landscape using classical continuation methods.
    }

\begin{figure}[htbp]
    \centering
    \includegraphics[width=0.7\textwidth]{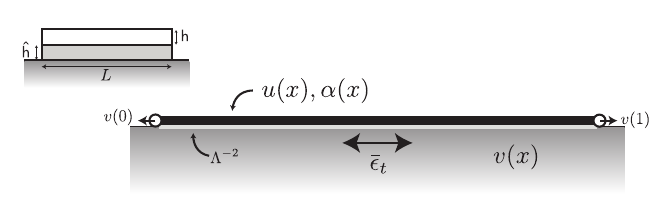}
    \caption{{Illustration of the thin film model (nondimensional). A thin film (black bar) is bonded to a compliant substrate (gray) via an elastic foundation (light gray) of effective stiffness $\Lambda^{-2}$, where $\Lambda$ is its characteristic length scale. The displacement $u(x)$ and damage variable $\alpha(x)$ are defined on the film, while $v(x)$ is the trace of the substrate's displacement at the interface. Boundary conditions prescribe the displacement $v(0), v(1)$, compatible with the time-parametrized average substrate strain $\bar{\epsilon}_t$. The inset shows the original 3D geometry before dimension reduction: a bilayer composed by a film of thickness $\hfilm$ bonded through an interfacial layer of thickness $\hinter$ over a domain of length $L$.}}
    \label{fig:thinfilm}
\end{figure}

\paragraph{Structural Model}
The structure is a multilayer composite:  a brittle thin film, whose material is identified by the state function $W$, bonded to a substrate which is either {stiff} or elastically compliant, {see Figure~\ref{fig:thinfilm}}. The thin film is a one-dimensional membrane with thickness $h$ and length $L$ with $L\gg h$, subject to a combination of imposed displacements by the substrate and loadings at the boundary (see Figure~\ref{fig:thinfilm}). The structure's reference configuration is the interval $(0, L)$.
The substrate is modelled as a one-dimensional elastic foundation whose displacement field is $v(x)$.
The displacement field which is elastically compatible to an homogeneous strain in the substrate is the linear function $v(x, t) = \bar\epsilon_t/2 (2x-1)$, where {$\bar\epsilon_t\in \mathbb R$} is the applied tensile {average} strain and $t>0$ plays the role of a loading time parameter.
The film is subjected to {the displacement of the substrate} $v(x, t)$ and to given (compatible) displacements at its free ends $x \in \{0, L\}$, so that for all $t$, $u(0)=v(0, t)$ and $u(L)=v(L, t)$
In the stiff, non-deformable substrate model, $\subsu$ is fixed. The elastic interaction is modelled by a distributed linear elastic foundation of stiffness $\mathsf{K}$, thus
the total energy of the structure is a functional $\Estiff$ \albdelete{constructed by considering the energy of the thin film and the energy associated with  the mechanical coupling between the film and the substrate. In nondimensional form, it is} given by

\begin{equation}
    \label{def:energy_stiff}
    \Psi( u, \alpha) = \int_{0}^{1} \left[ \frac{1}{2} \ \soften(\alpha)(u')^2
        + \frac{1}{2 \elastell^2} (u-\subsu)^2
        + \homogdiss(\alpha) + \frac{\damagell^2}{2}(\alpha')^2
        \right] dx,
\end{equation}
where $\Lambda^2 = \frac{\mathsf E_{\text{eff}}}{\mathsf{K}}$, $\mathsf E_{\text{eff}}$ being the effective stiffness of the two dimensional {membrane}. {Spatial variables and physical displacements are normalized with respect to the film's length $L$ and the displacement scale $u_0 := \frac{\mathsf{w_1}L}{\mathsf{K}}$}.
Our second model involves the same brittle thin film but a \emph{compliant} elastic substrate that undergoes deformation alongside the film. Unlike for the rigid substrate, the substrate's deformation is an additional unknown which affects the overall energy landscape of the system, incorporating an extra term accounting for the strain energy of the substrate.
The state of this structure is identified by the triplet $y:=(u, \alpha, v)$, and the nondimensional energy of the compliant system  $\Ecompl( u, \alpha, v)$ reads

\begin{equation}
    \label{def:energy_compliant}
    \Ecompl( u, \alpha,  v) = \int_{0}^1 \left[ \frac{1}{2} \ \soften(\alpha)(u')^2 + \homogdiss(\alpha) + \frac{\damagell^2}{2}(\alpha')^2
        + \frac{1}{2 \elastell^2} (u- v)^2
        + \frac{\stiffratio}{2}  (v')^2 \right] dx.
\end{equation} {The reduced elastic 1d model in~\eqref{def:energy_stiff} (and~\eqref{def:energy_compliant}) is deduced as the asymptotic vanishing thickness limit for a thin 3d elastic bilayer system of thickness  $\hfilm = o(L)$ {(see the inset in Figure~\ref{fig:thinfilm})} {where the thickness interface layer $\hinter=O(\hfilm)$ and its stiffness $\Youngsubs \ll \Youngfilm$ relate via dimensional scaling laws}, as shown in~\cite{Leon_Baldelli2015-rp}.
The nondimensional constants appearing in the expression of the reduced-dimension energies can be related to the three-dimensional parameters of the phyisical mechanical system}
\albdelete{Denoting by $\Youngfilm$ and $\hfilm$ the two-dimensional stiffness (per unit depth) and the thickness of the film membrane, and using a {hat} to indicate the stiffness and thickness of the substrate, the nondimensional quantities appearing in the expression of the energy depend on the geometric the elastic properties of the 3d system} as follows
\begin{equation}
    \label{def:dimensional_parameters}
    \damagell := \frac{\bar\damagell}{L}, \quad
    \Lambda := \frac{\Youngfilm}{\Youngsubs} {\frac{\hfilm \hinter}{L^2}}, \quad
    \stiffratio := \frac{\Youngsubs}{\Youngfilm} {\frac{\hinter}{\hfilm}}, \quad
\end{equation}
{where the hat indicates quantities related to the interface layer and the stiffnesses $\Youngfilm, \Youngsubs$ are per unit depth.}
Note that the nondimensional parameter $\Lambda$ relates the effective stiffness of the film membrane undergoing planar tensile strains (i.e. $\Youngfilm \hfilm$), to the effective stiffness of the bonding layer which, to accommodate the mismatch deformation between the film and the underlying substrate, is subject to a simple shear for which the effective stiffness is inversely proportional to its thickness, namely ${\mathsf{K}\sim}\Youngsubs / \hinter$.

For brevity, we denote by $y:=(u, \alpha)$ (and $y:=(u, \alpha, v)$) the mechanical state of the stiff (respectively, compliant) model system, and by $H^1(0, 1)$ the (Sobolev) space of scalar real functions defined on the unit interval which are square integrable and have square integrable first derivatives. The energy $\Estiff(y)$ is well-defined for pairs belonging in the Cartesian vector product space $V:=H^1(0,1)\times  H^1(0,1)$. Similarly, the energy $\Ecompl(y)$ is well-defined for triplets $(u, \alpha, \subsu)$ in $\widetilde V:=V\times  H^1(0,1)$.

\paragraph{Evolutionary model}

Assuming small enough loading rates,
the evolution problem for the structure can be cast \albdelete{in an energetic variational formulation} as an incremental rate-indipendent quasi-static process driven by an energy-minimality principle.
A state is observable only if it is stable, and in turn, a state is stable only if it is a local minimum of the energy among admissible state perturbations, at a given load level.

Differently from dissipative evolutions driven by an energy gradient flow whereby a system \emph{reaches} equilibrium conditions through a gradient descent process parametrized by an internal timescale, the quasi-static evolution we consider is a sequence of \emph{attained} equilibrium states, a necessary condition for local energy minimality.
The rate-indipendency further implies that the system does not exhibit internal timescales.
As such, energy minimisation is performed at any given value of the load, and at each increment of the external load its configuration evolves subject to imposed boundary conditions and possible internal constraints.
Specifically, we consider an evolution during the loading interval $t\in [0, T]$ as a time-parametrized mapping $t\mapsto y_t$  such that, for all $t\in [0, T]$ the realised (observed) state of the system $y_t$ is a local energy minimum among all admissible state perturbations. \albdelete{, or with respect to all admissible competitor states}%
In practice, given an initial condition $y_0$ at $t=0$ we seek a state $y_t:=(u, \alpha)_t$ (and $\widetilde{y_t}:=(u, \alpha, v)_t$, for the compliant model) such that, for a given value of the control parameter $t$, it satisfies time-dependent kinematic boundary conditions on the displacement variable and is {directionally stable}.
For definiteness, denoting by $X_t = \{v\in H^1(0, 1):v=\subsu(x, t),\text{ for } x=0\text{ and } x=1\}\times H^1(0,1)$ the affine vector space of (kinematically) admissible states for the stiff model, (respectively $\widetilde{X_t} = X_t\times H^1(0, 1)$, for the compliant model),
{the system at $y$ is directionally stable in the admissible direction $z$ if:}
\begin{equation}
    \exists  \bar{h} > 0  \text{ such that }  \Psi(y) \leq \Psi(y + h z), \quad \forall h \in [0, \bar{h}],
    \label{eq:variational_global_ineq}
\end{equation}
(and similarly for the compliant model substituting $\Estiff$ with $\Ecompl$ and $X_t$ with $\widetilde{X_t}$.) {
        Here $z$ is any admissible perturbation direction in the tangent space $T_y X_t$. If this condition is satisfied for all admissible directions $z \in T_y X_t$, the state $y$ is considered stable.
        Remark that this statement focuses on \emph{local} variations of the energy $\Psi$ around $y$, in line with numerical methods that explore the solution landscape directionally (e.g., quasi-Newton or Newton methods, even in finite-dimensional approximations), yet is defined through a \emph{global} quantity, the energy of the system $\Psi$.
        For small perturbations, this is a practical criterion that can be tested numerically without requiring a global search which is computationally prohibitive in {high}-dimensional spaces due to the vastness of the energetic landscape.
        In physical systems, stability often depends on small perturbations to the current state. Accordingly, this statement ensures that no small admissible perturbation $z$ decreases the energy of a stable state, which is a natural generalisation of a second-order stability condition.
        This
        condition is less restrictive than global energy minimisation but stricter than mere stationarity and is applicable for both reversible and irreversible systems since the admissibility of directions $z$ can incorporate \albdelete{irreversibility} constraints, ensuring that stability is tested only along physically meaningful directions.
        Compared to global minimisation, this approach is less restrictive, more realistic, in line with numerical methods, and provides a rigorous stability framework with a clear and physically meaningful \emph{global} condition for (local, directional) stability.
        In infinite-dimensional spaces, the notion of locality depends on the choice of topology (norm), which undermines the physical meaning of the solution, as the system's behavior should not depend on an arbitrary mathematical choice.
        The directional minimality statement above is appealing because it avoids the pitfalls of both global and norm-dependent local minimality, by leveraging the topology of the real line which is intrinsic. This aligns with physical intuition that systems tend to evolve along a \emph{path} according to stability and constraints.
    }

    {Remark that, in the current work, we do not require a global energy balance explicitly, as the quasistatic framework presented here does not resolve the jump transitions in sufficient detail to guarantee it.
        For smooth evolutions, it can be shown that an energy balance is satisfied at all times, as a consequence of the first order necessary conditions asociated to~\eqref{eq:variational_global_ineq}, cf.~\cite{pham:2011-gradient}, ensuring that the variation of (internal free) energy is compensated by external load. This budget structure does not hold globally for evolutions involving discontinuities.
        See~\cite{alessi:2016-energetic, larsen:2010}, for a detailed discussion of the energy balance in the context of variational evolutionary models.}

\paragraph{Admissibility of competitors and perturbations}
The admissibility of state competitors explicitly depends on the loading parameter through the kinematic boundary conditions (on displacement) and on internal constraints, namely whether damage (and hence the softening material behaviour)  evolves in a reversible or irreversible manner.
In the first case, as damage can evolve freely within the interval $[0, 1]$, all admissible states in $X_t$  (respectively, in $\widetilde{X_t}$) are also admissible competitors.
In the second case, the damage field is subject to the pointwise irreversibility constraint $\dot \alpha(x)\geq 0, \forall x\in [0, 1]$, requiring that the damage can only increase or stay constant.
As a consequence, irreversibility restricts the admissible set of competitors to the set $K^+_{\alpha_t}:=H^1(0, 1) \times \{\beta\in H^1(0, 1): \beta\geq \alpha_t\}$ (respectively, $\widetilde{K^+_{\alpha_t}}:=K^+_{\alpha_t}\times H^1(0, 1)$).
Remark that in the definition of the competitor space $\alpha_t$ (the damage field at time $t$) is unknown at time $t$.
In the irreversible case the set of admissible competitor states depends explicitly on the entire history of the evolution through the current damage field.
To draw the attention to the consequences of irreversibility on the system's transitions between different equilibrium states we develop the global variational inequality~\eqref{eq:variational_global_ineq} (and the analogous for the compliant model) by expanding the energy around the state $y_t$.
Admissible perturbations in the fully reversible case belong to the {tangent} space {$T_y X_t=X_0$} associated with  $X_t$ (respectively, ${T_y X_t=}\widetilde{X_0}$) {which is a linear vector space}, whereas in the irreversible case {admissible perturbations constitute} the closed convex cone $K^+_0$ (respectively, $\widetilde{K^+_0}$).

An energy expansion around $y_t$ reads
$$
    \Estiff(y)-\Estiff(y_t)= \delta\Estiff(y_t)(y-y_t)+\frac{1}{2}(y-y_t)^T \delta^2\Estiff(y_t)(y-y_t)+o(\|y-y_t\|^2),
    \label{eqn:energy-expansion}
$$
(and analogously for the compliant model), which allows to write first and second order (necessary and sufficient) conditions for optimality.

In the next section, we exploit the energy-stability inequality~\eqref{eq:variational_global_ineq} { to explore equilibria and identify transitions across branches.}

\section{Linear and nonlinear stability {of homogeneous solutions}}
\label{sec:stability}

Solutions to the incremental evolutionary problem are sought by solving first- and second- order necessary conditions for optimality encoded in the global variational inequality~\eqref{eq:variational_global_ineq}. This reduces to a \emph{linear} stability problem in a linear vector space for the case of fully reversible damage, and is a \emph{nonlinear} stability problem in a convex cone in the presence of irreversibility.

\paragraph{Linear stability in the reversible case - stiff substrate}

\begin{figure}
    \centering
    \begin{overpic}[width=\linewidth]{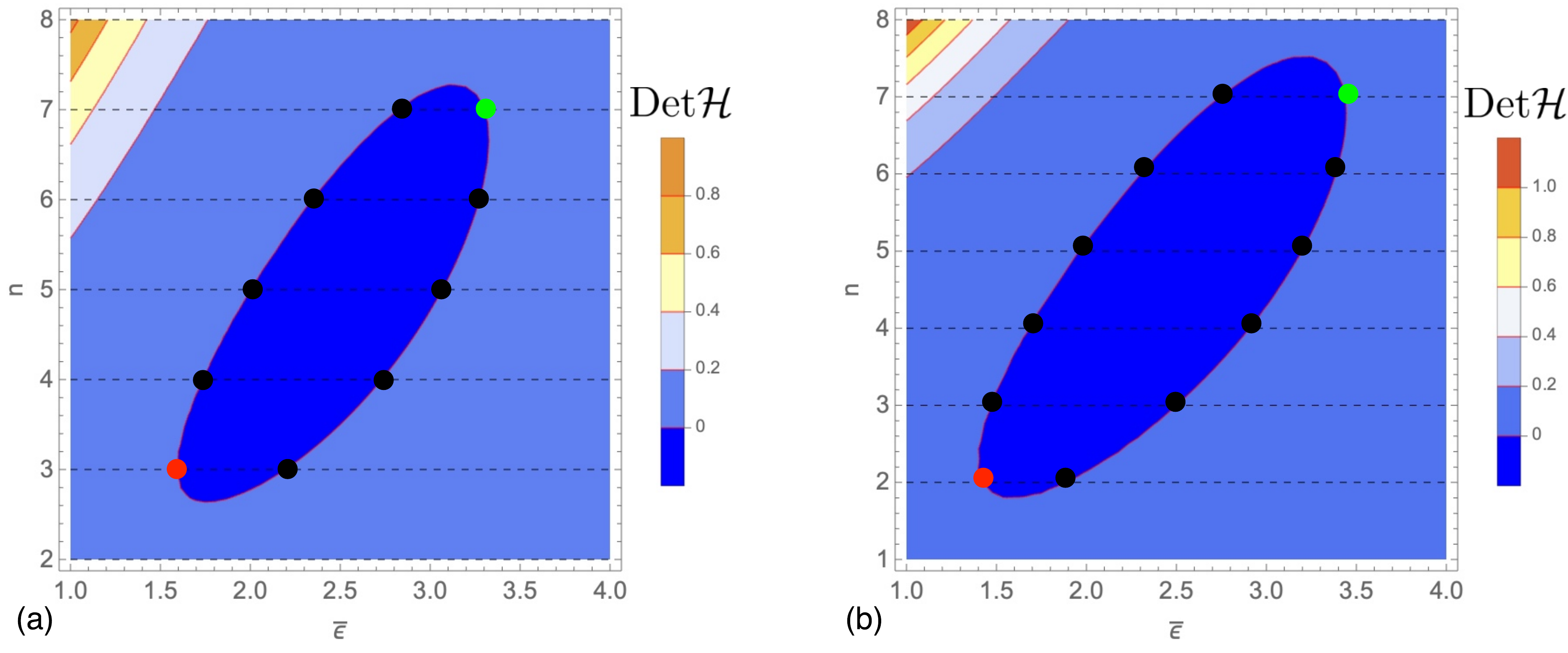}
        \put(33, 35){\textcolor{white}{$\bar \epsilon^{**}$}} %
        \put(88, 35){\textcolor{white}{$\bar \epsilon^{**}$}} %
        \put(59, 10){\textcolor{white}{$\bar \epsilon^*$}} %
        \put(7, 10){\textcolor{white}{$\bar \epsilon^*$}} %
    \end{overpic}
    \caption{{Determinant $\det H$ for homogeneous solution $\alpha_h$ ($\ell=0.16$, $\Lambda=0.34$) for: (a) rigid foundation; (b) compliant foundation ($\rho=0.5$). Closed loops indicate re-entry behavior at large deformations. Black dots: bifurcations from homogeneous solution; red/green dots: critical strains $\bar{\epsilon}^*$ and $\bar{\epsilon}^{**}$ (lower/upper stability limits). Critical wave numbers differ between limits: $n_c(\bar{\epsilon}^*) \neq n_c(\bar{\epsilon}^{**})$.}}
    \label{fig:hessian1}
\end{figure}

The first order variation of the energy functional $\Estiff$ in the direction $z:=(\utest, \beta)$ is given by the following linear form
\begin{equation}
    \delta \Estiff(u,\alpha)(\utest,\beta)=\int_0^1
    \left[\soften(\alpha)u'\utest'+\frac{1}{\elastell^2} (u-\subsu) \utest+ \left( \frac{1}{2}u'^2 \soften'(\alpha)+\homogdiss'(\alpha) \right)\beta+\damagell^2\alpha'\beta' \right]dx,\label{firstvar1}
\end{equation}
where
$z\in X_0$ is a test function (an admissible perturbation) in the linear space associated with  $X_t$. First order minimality conditions are \emph{local} conditions  associated with  the stationarity of the energy functional. The first order variation should vanish for all admissible test functions, namely
\begin{equation}
    \label{eq:stationarity}
    \delta \Estiff(u,\alpha)(w,\beta)=0, \quad \forall z{:=(w, \beta)}\in X_0.
\end{equation}
By using standard arguments of the calculus of variations, localizing the integral and choosing $\beta = 0$ first, and then $v =0$ leads to establishing the strong form \albdelete{of local stationarity conditions: the} mechanical equilibrium,  and the damage criterion. {Complemented by the associated boundary conditions,  they} are given by
\begin{eqnarray}
    \begin{cases}
        2(1-\alpha)\alpha' u' +(1-\alpha)^2 u'' -  \frac{1}{\elastell^2}(u-\subsu) & = 0, \quad {x\in (0, 1)} \\
        -\damagell^2\alpha'' - (1-\alpha)( u')^2 + 2\alpha                         & = 0, \quad {x\in (0, 1)}
    \end{cases}
    ,    \qquad
    \begin{cases}
        u(x)       = \subsu(x, t), & x=\{0, 1\}, \forall t \\
        \alpha'(x) = 0,            & x=\{0, 1\}, \forall t \\
    \end{cases}
    .
    \label{auto1}
\end{eqnarray}
Notice that the choice of boundary conditions for displacements compatible with the substrate's deformation implies that the pair $\yhom(\bar\epsilon):=(u_h(x), \alpha_h)(\bar\epsilon)$ given by $u_h(x)\equiv \subsu(x, t)$ and $\alpha_h$ a load-dependent constant to identify, is always (the unique homogenous) solution to the first order equilibrium equations. This makes it immediate to identify the fundamental homogeneous solution branch $t \mapsto \yhom$ and to decouple the elasticity problem from the  evolution of damage.
Therefore, the solution to~\eqref{auto1} such that $u''(x)=\alpha'(x)= 0, \, \forall x\in (0, 1)$ is the homogeneous branch
\begin{equation}
    u_h(x) =  \frac{\bar \epsilon_t}{2}(2x-1),\qquad\alpha_h = \frac{\Bar{\epsilon_t}^2}{2 + \Bar\epsilon_t^2}\label{eq:homo1}.
\end{equation}
Notice that,
first order conditions identify the critical load threshold that activates the damaging process. The quantity $\bar\epsilon_*^c$, the value of the load for which the undamaged elastic state becomes unstable, is determined injecting the elastic solution $u_h$ in the energy and computing $\delta \Estiff(u_h, 0)(0, \beta)=0$.
According to our mechanical energy model, the critical load is given by
\begin{equation}
    {{{\bar{\epsilon}^c}_*}} = \sqrt{2\frac{\homogdiss'(0)}{\soften'(0)}}=0,
\end{equation}
{due to the quadratic $\homogdiss(\alpha)$.}
The effective {total} energy along the {homogeneous} branch reads
\begin{equation}
    \label{eq:energy_homogeneous}
    \Estiffhom(\Bar{\epsilon_t}) :=\Estiff(\yhom(\Bar{\epsilon_t})) = \frac{\Bar{\epsilon_t}^2}{2 + \Bar{\epsilon_t}^2}.
\end{equation}
An equilibrium configuration $y_t:=(u,\alpha)_t$ is a state such that the first variation $\delta \Estiff(y_t)(z)$ vanishes for all admissible test fields in the vector space $X_0$.
To assess the incremental stability of the homogeneous solution in the reversible (linear) case, we examine the positivity of the second variation, requiring
\begin{equation}
    \delta^2 \Estiff(y_t)(y-y_t, y-y_t)>0, \qquad  \forall {y-y_t}\in X_0,
    \label{eqn:linear_second_order_stability}
\end{equation}
{The second directional derivative of the energy is given by the following bilinear form}
\begin{equation}
    \delta^2 \Estiff(u,\alpha)({\utest},\beta)=\int_0^1 \left[(1-\alpha)^2{\utest}'^2
    +\frac{1}{\elastell^2} {\utest}^2 \right]dx
    +\int_0^1
    \left[ - 4(1-\alpha)u' {\utest}'\beta+(2+ u'^2)\beta^2+\damagell^2\beta'^2 \right] dx,
    \label{hessian22}
\end{equation}
which is well defined for perturbations $(\utest, \beta)\in X_0$.
    {
        We work with fields in the Sobolev space $H^1(0,1)$, where the sine/cosine Fourier basis (adapted to the boundary conditions) forms a dense subset. Any admissible perturbation can thus be approximated arbitrarily well in the  energy norm by a finite Fourier sum. The spectral method we employ therefore provides a sufficient condition for instability (if at least one eigenvalue is negative).
        Moreover, since the spectrum of the second variation operator is discrete, bounded below, and diverges to infinity, the lowest eigenvalue is attained at a finite mode index $n$ (typically among the lowest modes) thus ensuring a sufficient condition for stability as well.
    }
    {We} hence extract information on the onset of instability by seeking a solution in Fourier series of the fields $\utest$ and $\beta$ in \eqref{hessian22}, such that $$\utest(x)=\sum_{n=1}^{\infty} a_{n} \sin \left(n \pi x+\phi_{n}\right), \quad \beta(x)=\sum_{n=1}^{\infty} b_{n} \cos \left(n \pi x+\psi_{n}\right).$$
Then, by boundary conditions, $\psi_{n}=\phi_{n}=0$ for all natural $n$. The stability condition~\eqref{eqn:linear_second_order_stability} {for homogeneous states $\yhom$} takes the form:
\begin{align}\left[ a_n \quad b_n \right] \mathcal{H} \left[ \begin{array}{c} a_n \\ b_n \end{array} \right]=\left[ a_n \quad b_n \right]\left(
    \begin{array}{cc}
            \soften(\alpha)(n\pi)^2+\elastell^{-2} & \soften{'(\alpha)}\bar\epsilon n\pi \\
            \text{sym.}                            & \soften{''(\alpha)}\bar\epsilon^2
            + {\homogdiss''(\alpha)}+\damagell^2(n\pi)^2                                 \\
        \end{array}
    \right)\left[ \begin{array}{c} a_n \\ b_n \end{array} \right]>0.\label{hessian1}\end{align}
{The trace of $\mathcal{H}$ is always positive along the homogeneous path due to the monotonicity of the underlying quadratic terms, hence the sign of $\det \mathcal{H}$ alone governs the stability transition.}
By substituting the homogeneous solution $(u_h, \alpha_h)$ into \eqref{hessian1}, we compute $\det \mathcal{H}$ as a function of $\bar \epsilon$ and $n$, see Figure~\ref{fig:hessian1}. The calculations are performed for the parameter values $\damagell=0.16$ and $\elastell=0.34$.
The figure represents, for a given load $\bar \epsilon$, the wave number $n(\bar \epsilon)\in \mathbb N$ of possible energy-decreasing damage bifurcations. For an increasing loading history $\bar \epsilon_t\nearrow$, the wave number is non-monotonic.
In Fig. \ref{fig:hessian1}(a), the locus $\det \mathcal H=0$ forms closed loops, indicative of an elastic background's influence. Notably, {observe} the re-entry behavior of the affine configuration  (i.e., the homogeneous configuration re-stabilizes at large deformation). This is marked by the emergence of two critical strains denoted as $\bar\epsilon^*$ and $\bar\epsilon^{**}$ (with $\bar\epsilon^* < \bar\epsilon^{**}$), representing the lower and upper stability limits for the homogeneous state. These critical points are highlighted by red and green dots in Fig. \ref{fig:hessian1}(a). The critical wave number $n_c$ for the lower limit $n_c(\bar{\epsilon}^*)$    differs from the critical wave number for  upper limit $n_c(\bar{\epsilon}^{**})$. Finally, we remark that closed-form analytical solutions can be provided for the critical wave number and critical strains and the parametric dependence of the corresponding bifurcation thresholds can be obtained, as detailed in \cite{Salman2021-mn}.

\paragraph{Linear stability in the reversible case - compliant substrate}

Denoting by $z:=(\utest, \beta, \subsutest)\in \widetilde{X}_0$ a test function for the state triplet  $y_t:=(u, \alpha, \subsu)_t$ at time $t$, the first order variation of the energy functional $\Ecompl$ is given by the following linear form
\begin{equation}
    \delta \Ecompl(u, \alpha, \subsu)(\utest,\beta,\subsutest)=\int_0^1 [(1-\alpha)^2u'v'+\frac{1}{\elastell^2} (u- v) (v- \subsutest)+\stiffratio \subsu'\subsutest'+\frac{1}{2}u'^2 (\soften'(\alpha)+\homogdiss'(\alpha))\beta+\damagell^2\alpha'\beta' ]dx,\label{firstvar}
\end{equation}

Arguing similarly to the stiff case by localization and integration by parts, for all $x \in (0, 1)$ the Euler-Lagrange equations and associated boundary conditions read
\begin{eqnarray}
    \label{modeld_el_1}
    \begin{cases}
        2(1-\alpha)\alpha' u' +(1-\alpha)^2 u'' -  \frac{1}{\elastell^2}(u-\subsu) & = 0, \\
        -\damagell^2\alpha'' - (1-\alpha)( u')^2 + 2\alpha                         & = 0, \\
        \stiffratio  \subsu''  +  \frac{1}{\elastell^2}(u-\subsu)                  & = 0, \\
    \end{cases}
    ,    \qquad
    \begin{cases}
        u(x)       = \subsu(x, t), & x\in\{0, 1\}, \forall t \\
        \alpha'(x) = 0,            & x\in\{0, 1\}, \forall t \\
    \end{cases}
    .
    \label{auto2}
\end{eqnarray}
It is easy to show that the homogeneous solution  remains the same as in the case of the rigid substrate, that is $\alpha_h(\bar{\epsilon}) = \frac{\bar{\epsilon}^2}{2 + \bar{\epsilon}^2}\label{eq:homo11}$, whereas the effective elastic energy along the homogeneous branch now reads $\Ecomplhom(\bar{\epsilon}) = \frac{\bar{\epsilon}^2}{2 + \bar{\epsilon}^2} + \frac{\rho}{2}\bar{\epsilon}^2$.

We once more seek the linear incremental stability of an equilibrium configuration $y_t := (u, \alpha,  v)_t$ satisfying that the first order condition $\delta \Ecompl(y_t)(y-y_t)=0$ for all admissible state perturbations $z:=y-y_t\in\widetilde X_0$ by examining the positivity of the second variation, namely
\begin{equation*}
    \delta^2 \Ecompl(y_t)(y-y_t, y-y_t)>0, \qquad \forall y-y_t\in \widetilde X_0,
\end{equation*}The second variation is given by the  bilinear form
\begin{equation}
    \delta^2 \Ecompl(y_t)(y-y_t, y-y_t)=\int_0^1 \left[(1-\alpha_t)^2\utest'^2
        - 4(1-\alpha_t)u_t' \utest'\beta+(2+ {u_t'}^2)\beta^2+\damagell^2\beta'^2 +\frac{1}{\elastell^2}( \utest^2 + \subsutest^2)+\stiffratio \subsutest'^2\right]dx.\label{hessian222}\end{equation}
We again proceed to extract information on the onset of instability expanding in Fourier series the fields $\utest$, $\beta$ and $\subsutest$  in \eqref{hessian222} such that
\[
    \utest(x) = \sum_{n=1}^{\infty} a_{n} \sin \left(n \pi x + \phi_{n}\right),
    \quad\beta(x) = \sum_{n=1}^{\infty} b_{n} \cos \left(n \pi x + \psi_{n}\right),
    \quad\subsutest(x) = \sum_{n=1}^{\infty} c_{n} \sin \left(n \pi x + \theta_{n}\right).
\]
Similar to the previous section we claim that, whenever the first order term vanishes, the system is stable only if $\delta^2 \Ecompl
    >0$ for all sufficiently smooth admissible test fields $(\utest, \beta, \subsutest)$ in the vector space $\widetilde{ X_0}$. The stability condition {for the homogeneous state $\yhom$}  takes the form
\begin{align}
    \left[ a_n \quad b_n \quad c_n \right]\left(
    \begin{array}{ccc}
            \soften(\alpha)(n\pi)^2+\elastell^{-2} & n\pi {\soften'(\alpha)}\bar\epsilon                                          & -\elastell^{-2}                      \\
            {\text{sym.}}                          & {\soften''(\alpha)}\bar\epsilon^2+ {\homogdiss''(\alpha)}+(\damagell n\pi)^2 & 0                                    \\
            {\text{sym.}}                          & {\text{sym.}}                                                                & \elastell^{-2} +\stiffratio (n\pi)^2 \\
        \end{array}
    \right)\left[ \begin{array}{c} a_n \\ b_n \\c_n \end{array} \right]>0.\label{hessian3}\end{align}
By substituting the homogeneous solution {$\alpha_h$} into \eqref{hessian3}, we compute $\det \mathcal{H}$, see Fig. \ref{fig:hessian1}(b). The calculations are performed for the parameter values $\damagell=0.16$, $\elastell=0.34$ and $\rho=0.5$.  In Fig. \ref{fig:hessian1}(b), we again observe the formation of closed loops and a re-entry behavior of the affine configuration is discernible. This is marked by the emergence of two critical strains denoted as $\bar\epsilon^*$ and $\bar\epsilon^{**}$, representing the lower and upper stability limits for the homogeneous state, highlighted by red and green dots in Fig. \ref{fig:hessian1}(b). The critical wave number $n_c$ for the lower limit $n_c(\bar{\epsilon}^*)$    differs from the critical wave number for the upper limit $n_c(\bar{\epsilon}^{**})$. The overall behavior of the system remains the same while the first critical wave number $n_c=2$ is now smaller suggesting a different number of cracks will appear during  loading.

\paragraph{Nonlinear stability with irreversibility constraints}

Irreversibility is introduced in the variational formulation of the evolution problem as a pointwise inequality constraint. It plays two distinct roles along an evolution: it acts i) as a \emph{local} constraint which prevents the damage field at a given location to decrease between two subsequent loads values (both during monotonic and non-monotonic load programs), and ii) as a global restriction of the space of admissible variations in such a way that \emph{negative} perturbations of the current damage state are no longer allowed. This changes the topological structure of the set of admissible perturbations, from a linear vector space to a convex cone.
To enforce irreversibility we consider only non-decreasing damage evolutions that are sufficiently smooth with respect to the loading parameter, and seek maps $t\mapsto y_t = (u_t, \alpha_t)$ such that $\dot \alpha_t \geq 0$,  satisfying the minimality condition~\eqref{eq:variational_global_ineq}.
Because the current state can only be compared to those of with equal or higher damage, the space of admissible perturbations is a  convex cone strictly contained in the vector space of admissible unconstrained perturbations, namely $K^+_0\subset X_0$. Indeed, for $(v, \beta)\in K^+_0$, then $-(v, \beta) \in X_0$ but $-(v, \beta)\notin K^+_0$. This restriction has a profound impact on the variational characterization of local minima and bears consequences on both  the first order (equilibrium) conditions and the second order (stability) problem which become \emph{unilateral} conditions.
Equilibrium states $y_t=(u_t, \alpha_t)$ of the irreversible system are hence governed by the following (first order) necessary optimality conditions taking the form of a variational inequality
\begin{equation}
    \label{eq:eq_variational_inequality_full}
    \delta \Estiff(y_t)(y-y_t) \geq 0, \quad \forall y-y_t \in K^+_0.
\end{equation}
This must hold for all admissible competitor states $y$ such that $v-u_t \in H^1_0(0, 1)$ and $\beta-\alpha_t\in H^1(0,1)$ with $\beta \geq \alpha_t$.
Testing separately the elasticity (for fixed damage), and the damage problems (for a given displacement), we obtain
\begin{equation}
    \label{eq:variational_equilibrium}
    \delta \Estiff(y_t)(v-u_t, 0) = 0, \qquad \delta \Estiff(y_t)(0, \beta-\alpha_t) \geq 0, \qquad \forall y-y_t \in K^+_{0}.
\end{equation}
The last two relations are, respectively, the weak form of the mechanical equilibrium conditions and the  evolution law  for the damage field.
The former {yields} the elastic equilibrium $u_t$ which can thus be eliminated from energy, while the latter governs the evolution of the damage field.
Upon elimination of the kinematic field, the variational inequality~\eqref{eq:variational_equilibrium}$_2$ takes a particularly expressive form when written as a complementarity problem. This highlights the mechanical nature of the damage criterion as a threshold law.
Namely, the strict convexity of the elastic model for given damage implies that~\eqref{eq:variational_equilibrium}$_1$ has, for given $\alpha$, a unique time-parametrized solution $u_t(\alpha)$.
Substituting in ~\eqref{eq:variational_equilibrium}$_2$ and accounting for the irreversibility constraint we are led to seek a {sufficiently smooth (continuous)} map $t\mapsto\alpha_t$ such that
\begin{equation}
    \label{eq:complementarity}
    \dot \alpha_t \geq 0 \qquad
    -\phi_t(\alpha_t) \leq 0 \qquad
    \phi_t(\alpha_t)\dot \alpha_t = 0.
\end{equation}
{Note that the complementarity conditions above hold during intervals of smooth (time-continuous) evolution; in particular, \eqref{eq:complementarity}$_3$ can be integrated as an energy budget, valid for all intervals where $\alpha_t$ evolves continuously but not across jump discontinuities.}
In \eqref{eq:complementarity}, $\phi_t$ is the scalar function associated with  the variation of elastic energy density {computed} at the equilibrium {and} defined by $\delta \Estiff(y_t)(0, \beta) = \langle -\phi_t(\alpha_t), \beta\rangle$.
Consequently, $\phi_t(0)$ is
the variation of the  energy at equilibrium for the undamaged structure. All equilibrium solutions $u_t$ such that $-\phi_t(0) > 0$ belong to the interior of the damage yield surface.
The equality $-\phi_t(0) = 0$, conversely, identifies the elastic limit and corresponds to the damage criterion for the {(undamaged, $\alpha=0$)} structure. {Equivalently, attainment of the equality means} that the state $(u_t, 0)$ has reached, from the interior, the boundary of the (damage-dependent) elastic domain.
Explicitly, $\phi_t$ {is given by}
\begin{equation}
    \label{eq:energy_release_rate}
    \phi(\alpha) := \frac{1}{2}\soften'(\alpha){e}^2 + \homogdiss'(\alpha){+ \ell^2 \alpha''(x)},
\end{equation}
where ${e}$ is the elastic strain and $\soften'(\alpha)$ and $\homogdiss'(\alpha)$ are the derivatives of the softening and dissipation energy densities with respect to the damage variable.
\albdelete{The inequality ~\eqref{eq:variational_equilibrium}$_2$ identifies the domains of admissible strains (and, by duality, of stresses) for homogeneous solutions as
    $\mathcal{R}(\alpha):=\{e\in \mathbb R^{2\times 2}_{sym}: \Youngfilm e^2 \leq -\frac{2 \mathsf{w}'(\alpha)}{\mathsf{a}'(\alpha)}\},$ and $
        \mathcal{R}^*(\alpha):=\{\sigma\in \mathbb R^{2\times 2}_{sym}: \frac{\sigma^2}{\Youngfilm} \leq \frac{2 \mathsf{w}'(\alpha)}{\mathsf{s}'(\alpha)}\}$, respectively.}
As a first order optimality condition,~\eqref{eq:variational_equilibrium}$_2$ states that the local elastic energy release is either smaller than or equal to the (marginal) cost of damage, whereas the complementarity condition $\phi_t(\alpha_t)\dot \alpha_t = 0$ ensures that the damage field evolves only if the energy release rate is critical.
The three conditions above
    {constitute the set of first order Karush-Kuhn-Tucker conditions for the complementarity problem~\cite{Nocedal1999-zr}}.

{Assuming} homogeneous initial conditions $y_0=(0, 0)$ and compatible kinematic boundary conditions, the existence of a homogeneous solution {(such that for all  $x\in (0,1), u''(x)=\alpha'(x)=0$)} implies that {at a given load level} the damage criterion is attained everywhere throughout the bar\albdelete{ at the same load}. This greatly simplifies the analysis of the energetic properties of the system.
Using the elastic solution $u_t = 2\bar \epsilon_t(x-1/2)$
the inequality in~\eqref{eq:variational_equilibrium}$_2$ yields the following algebraic inequality
\begin{equation}
    \label{eq:variational_equilibrium_homogeneous}
    0 \geq -\phi_t(\alpha_t)= -\frac{1}{2}\bar \epsilon_t^2 \soften'(\alpha_t)-\homogdiss'(\alpha_t),
\end{equation}
which identifies the evolution of the homogeneous damage response (${\alpha''(x)}$), as a function of the given load level $t$.
The investigation of the stability properties requires
considering second order energy variations with respect to all admissible perturbations nullifying the first order term in expansion~\eqref{eqn:energy-expansion}.
\albdelete{In the general case, this requires distinguishing between the regions where the damage criterion is attained, and thus damage can evolve, from the (complementary) domain where damage cannot evolve (there, the second relation in~\eqref{eq:complementarity} is satisfied with a strict inequality).
    In our setup, the existence of nontrivial homogeneous solutions simplifies the analysis because the damage criterion is attained everywhere thus  the function space of admissible perturbations is defined on the fixed domain $(0, 1)$.}

Assume now that a state $y_t$ is known as a function of $t$ such that it solves~\eqref{eq:eq_variational_inequality_full} and is sufficienty smooth so that the (right) derivative with respect to $t$ is well-defined.
As $t$ varies, $y_t$ describes a (smooth) curve in the phase space identified by its right tangent vector $\dot y_t =: \lim_{\tau\to 0^+}\frac{y_{t+\tau} - y_t}{\tau}$, the rate of evolution.
A key question is whether $y_t$ traces a unique evolution path, or conversely if it lays at the intersection of multiple equilibrium curves.
To this end, {under the assumption that $t \mapsto y_t$ is differentiable and (formally) applying the chain rule, differentiating~\eqref{eq:eq_variational_inequality_full} we have}
\begin{equation}
    \frac{d}{dt} \left[\delta \Psi(y_t)(\zeta - y_t) \right]= \delta^2 \Psi(y_t)(\dot{y}_t, \zeta - y_t) + \delta \dot\Psi(y_t)({\zeta} - {y}_t) - \delta \Psi(y_t)(\dot y_t),
\end{equation}
where $\delta \dot\Psi(y_t)$ is the partial time-derivative of the linear form corresponding to the first order energy variation.
    {Using the first-order condition again, the last term vanishes.}
    {We thus obtain a rate form of the variational inequality} relating the rate of evolution $\dot y_t$ to the current state $y_t$ (supposing the latter known {and smooth}), namely
\begin{equation}
    \label{eq:variational_bifurcation}
    \qquad \delta^2 \Psi(y_t)(\dot y_t,   \zeta - y_t) + \delta \dot \Psi(y_t)(\zeta- y_t) \geq 0, \quad \forall \zeta\in X_{0},
\end{equation}
{
Whenever the energy does not explicitly depend on $t$, the second (linear) term vanishes.
Above, the quantity $\dot{y}_t$ is the tangent direction to the evolution curve, i.e., the rate at which the system evolves under a slow increase in loading.
Physically, it is a candidate (direction for the) evolution path: if the system is at a regular point (not at a branching point), then $\dot{y}_t$ is unique. Contrarily, if $\dot{y}_t$ lies in a flat or saddle region of the energy landscape, multiple directions of motion are possible, and bifurcation (branching of solution paths) can occur.
Thus, if \eqref{eq:variational_bifurcation} admits a unique solution $\dot{y}_t$, the evolution path is unique; otherwise, multiple trajectories may bifurcate from $y_t$.}
By construction,
the homogeneous rate $\dot y_h$ is a solution of \eqref{eq:variational_bifurcation} and the question is whether another solution exists.
The uniqueness of the homogeneous evolution is thus ensured by the positive definiteness of the second variation over $X_0$, so that the non-bifurcation condition reads
\begin{equation}
    \label{eq:bifurcation_uniqueness}
    \delta^2 \Psi(\yhom)(\zeta, \zeta) > 0, \quad \forall \zeta \in {X_0\setminus \{0\}},
\end{equation}
which formally coincides with the classical \emph{linear stability} problem~\eqref{eqn:linear_second_order_stability} in the reversible case, yet has a different mechanical interpretation, here, in terms of the uniqueness of evolution rates and the absence of branching equilibrium paths.
\albdelete{Remark that, in the general case in which the damage criterion is not attained everywhere, the space of admissible perturbations for the (second order) bifurcation problem is $X_0^\dagger:=H^1_0(0, 1) \times \{ \beta \in H^1(0, 1) : \delta\Psi(y_t)(0, \beta) = 0 \}$ for the stiff substrate model, and $\widetilde{X_0^\dagger}:=X_0^\dagger\times H^1_0(0,1)$ for the compliant substrate model.
}
The first  bifurcation load is $t_b:=\inf_t \{\delta^2 \Psi(y_h)(\zeta, \zeta) =0, \forall \zeta \in X_0{\setminus \{0\}} \}$. {the infimum load where the second variation~\eqref{eq:bifurcation_uniqueness} ceases to be positive definite, thus} there exist (multiple) equilibrium curves intersecting the homogeneous branch.
Thus, for $t\geq t_b$ bifurcating away from homogeneous branch becomes possible.
The study of the bifurcation problem is functional to infer a partial response on the stability of the state. Indeed, if the current equilibrium branch is unique then, necessarily, the current state is stable. The converse is not true {the irreversible case}, as the existence bifurcation paths is not a sufficient condition to rule out the stability of the current state.

The stability of the homogeneous solution in the irreversible case, according to our energetic viewpoint, is governed by the positivity of $\delta^2 \Psi(\yhom)$ on the constrained space of admissible state perturbations $K^+_0$.
Defining $t_s: = \inf_t \{\delta^2 \Psi(\yhom)(\zeta, \zeta) =0, \forall \zeta \in K^+_0 \}$ the load at which the homogeneous solution loses stability, the set inclusion $K^+_0 \subset X_0$ implies that necessarily $t_b \leq t_s$.
Equality occurs when the first bifurcation mode {has a definite sign}.

This indicates a qualitative conceptual distinction between the bifurcation and the stability thresholds, in the irreversible case. As a consequence, a system can persist along a critical non-unique equilibrium branch, yet be stable.
A sufficient condition for the stability of the homogeneous state $\yhom$ is given by the strict positivity  of the Hessian form on the constrained space of admissible perturbations, namely (for the stiff substrate model)

\begin{equation}
    \label{eq:variational_stability}
    \delta^2 \Psi(\yhom)(y - \yhom,  y - \yhom)  > 0, \quad \forall y-\yhom \in \conespace,
\end{equation}
and similarly for the compliant substrate model, by replacing $\Estiff$ with $\Ecompl$ and $\conespace$ with $\widetilde{\conespace}$.
The variational inequality above is a constrained eigenvalue problem {that} characterizes the stability of the state $\yhom$. Its solution yields, at load $\epsilon_t$, either a positive eigenpair $(\lambda_t, z^*_t)\in \mathbb{R}^+\times K^+_0$ as a sufficient condition for the stability of current state, or a pair $(\lambda_t, z^*_t)\in \mathbb{R}^-\times K^+_0$ where $\lambda_t$ is the local (negative) energy curvature and $z^*_t$ the eigenmode, indicating the direction of maximum energy decrease. This is interpreted as the \emph{instability mode}, pointing the system towards an optimal direction of energy descent.
    {the (linear) eigenproblem `in the ball' \eqref{eq:variational_bifurcation} enables bifurcation analysis, with admissible perturbations lie in a linear norm-bounded vector space.
        By contrast, the (nonlinear) `cone-constrained' eigenproblem \eqref{eq:variational_stability} captures stability under inequality constraints (irreversibility), by restricting variations to a convex cone (e.g., non-negative variations in damage).
    }
From the numerical standpoint, the bifurcation eigenproblem in the vector space~\eqref{eq:variational_bifurcation} may be regarded as an approximated version of the stability problem, in the time-discrete setting.
Indeed, as suggested in~\cite{Baldelli2021-gc} through the notion of `incremental-stability', the irreversibility constraint in the stability problem can be relaxed to a pointwise inequality with respect to the state at the previous time-step.
\albdelete{, denoted $y_-$.
    The mechanical intuition is to replace the \emph{current} state $y_t$ with $y_-$, in the definition of the space of perturbations.
    As a consequence, denoting by $\alpha_t$ the equilibrium damage field solving first order optimality conditions at $t$ and by $\alpha_-$ is the solution at the previous load step, admissible perturbations for the second order problem~\eqref{eq:variational_stability} are all the $y-y_-\in \conespace$. In this way, the set of perturbations is enlarged. It includes all sufficiently smooth functions $\beta$ which cancel the first order term, without restriction on the sign provided that $\alpha_t(x) + \beta(x) - \alpha_-(x)\geq 0$ for all $x\in (0, 1)$.
    Consequently, the space of perturbation allows for (small) perturbations $\beta$ which can be negative at points $x\in(0,1)$ where $\alpha_t(x)>\alpha_-(x)$.}
Such a space is a vector space and the eigenproblem can be solved by standard methods of linear algebra by projecting the Hessian to the set of active constraints, cf~\cite{Nocedal1999-zr}.
Conversely, the stability problem~\eqref{eq:variational_stability} is of a different nature due to the different topology of the underlying space of variations, which requires specialised numerical methods.

\section{Numerical solutions}
\label{sec:numerics}
\subsection{Identification of equilibrium branches {for the reversible case}}

We now turn our attention to the solutions beyond the  homogeneous branch to identify all the equilibrium branches corresponding to the inhomogeneous {localised} solutions. Our goal is to construct an \emph{equilibrium map} that represents all stable and unstable equilibrium states as a function of the external load~\cite{Pattamatta2014-pn}.
To do so, we use the pseudo-arclength continuation technique implemented in the software AUTO~\cite{Doedel1981-sa}, see also \cite{Pattamatta2014-pn}.
It uses collocation with Lagrange polynomials to discretise the boundary-value problem, in our simulations we had $300$ mesh points with $5$ collocation nodes and activated mesh adaptation.
    {Then, } it solves the nonlinear equations \ref{auto1} (and \ref{auto2} in the compliant case), with the {end displacement} $\bar\epsilon_t$ treated as a continuation parameter.

\begin{figure}
    \centering
    \hspace*{-.3cm}
    \includegraphics[align=c, width=.6\textwidth]{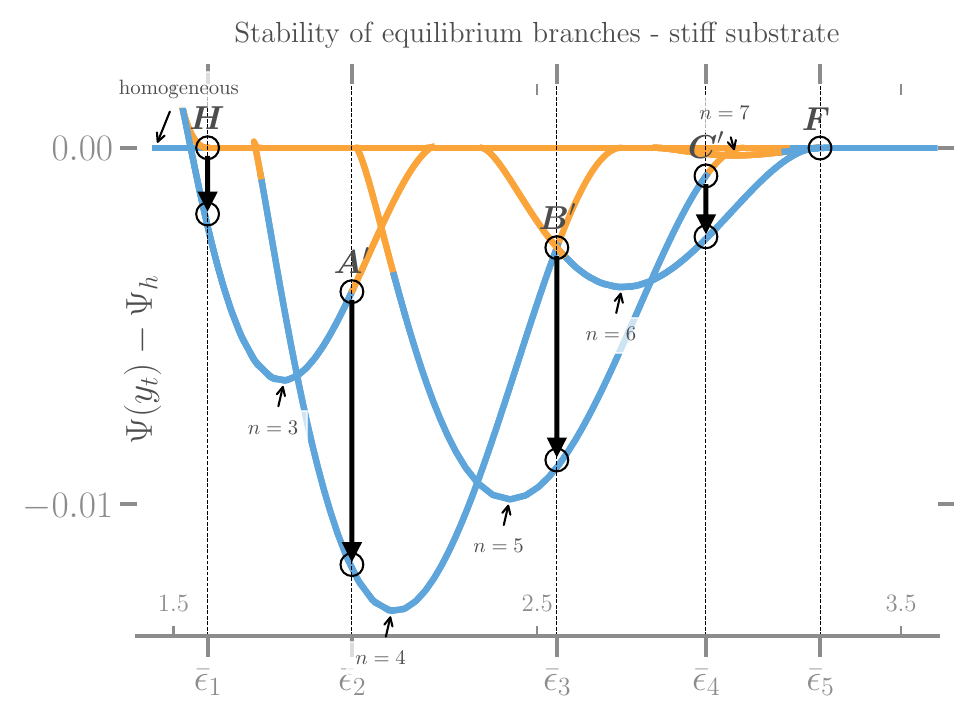}
    \includegraphics[align=c, width=.4\textwidth]{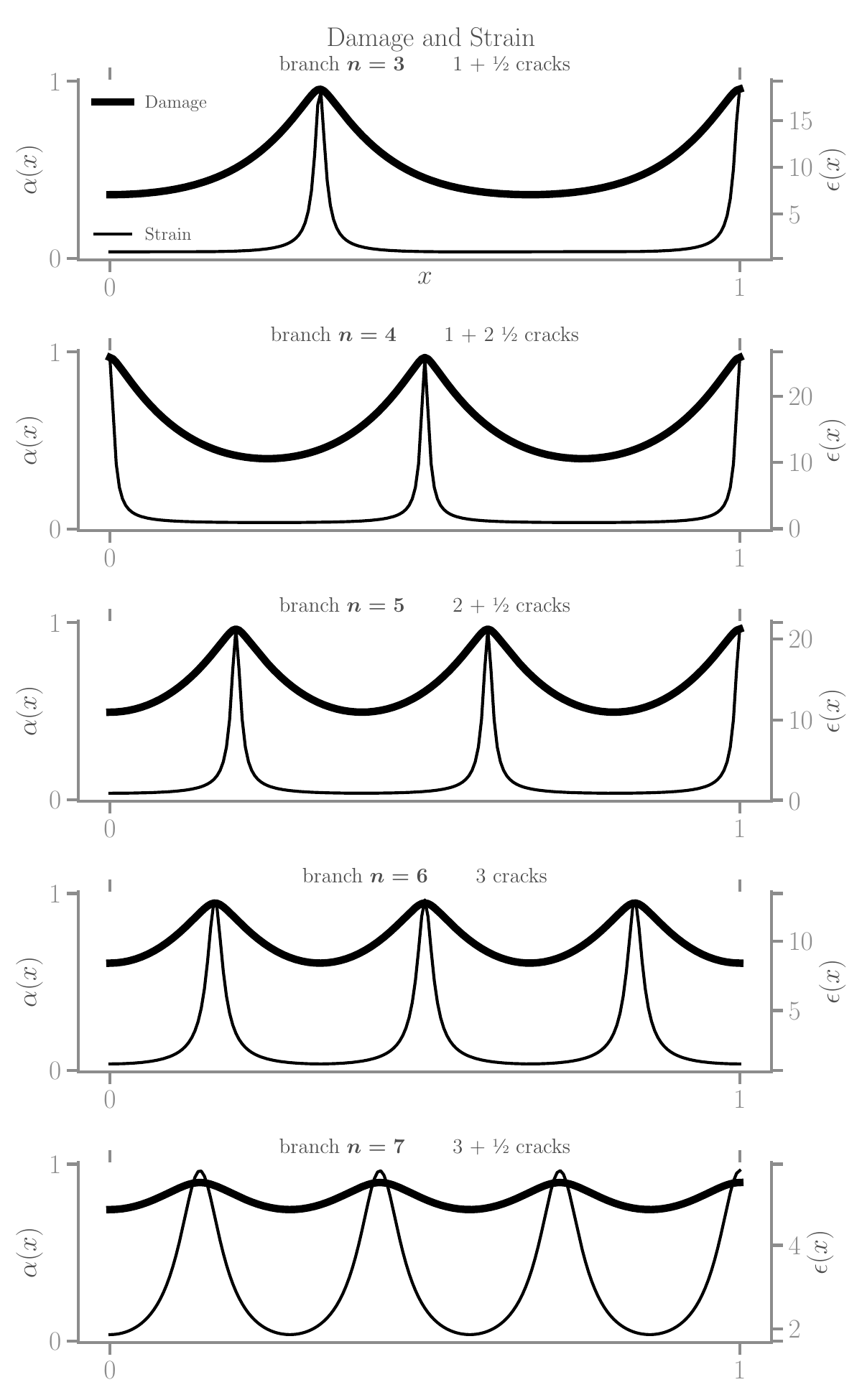}
    \includegraphics[align=c,width=.8\textwidth]{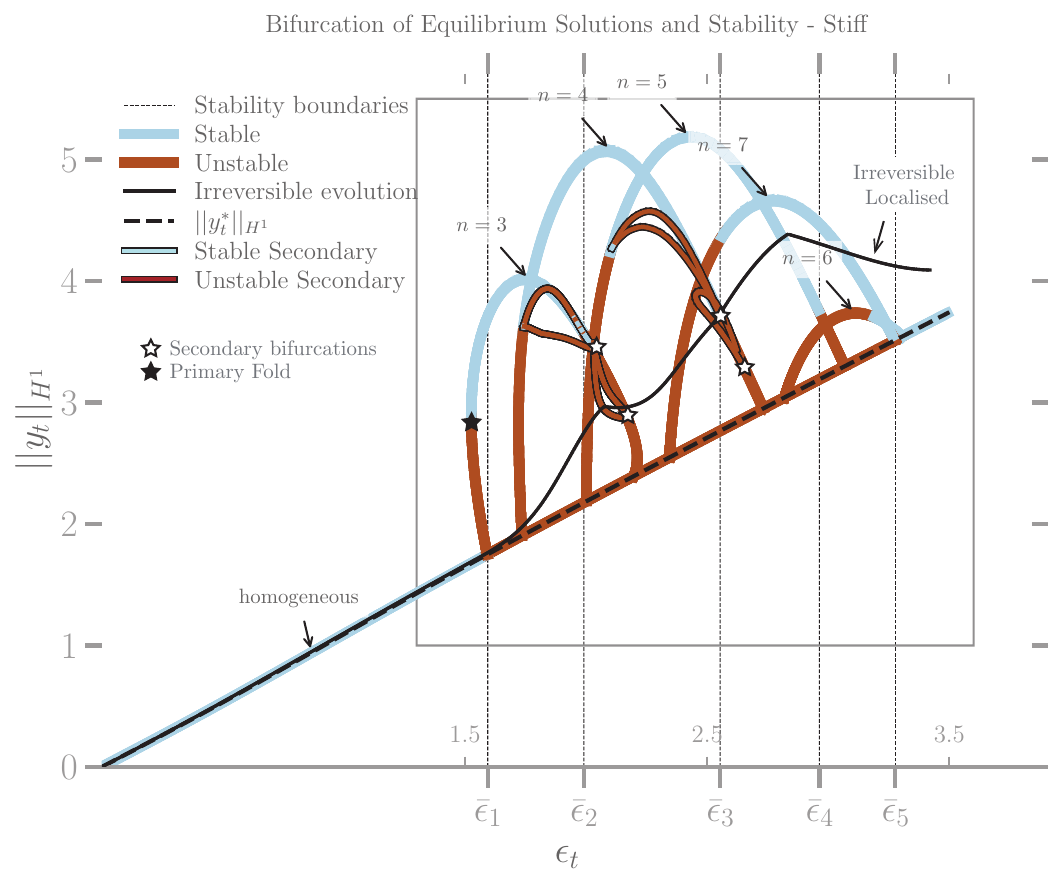}
    \caption{{Equilibrium branches for phase-field thin film model on rigid elastic foundation. Left: energy difference $\Delta\Psi$ between current and homogeneous configurations at given load. Stability is color-coded: blue (stable), orange (unstable) based on smallest eigenvalue of stiffness matrix $H$. Arrows indicate branch switching at stability loss. Branches parametrised by integer $n$ (cf. profiles); Right: damage and strain profiles of minimum energy configurations on each branch. Bottom: bifurcation and stability map. Solution curves starting from the homogeneous zero-damage state. The first bifurcation along the homogeneous branch is a subcritical stability-switching point, followed by a primary fold where stability is regained. Secondary bifurcation points along localized branches are also indicated; note that some bifurcations occur without affecting stability. The irreversible evolution is shown in black (dashed line for homogeneous solutions, solid line for localized ones); both remain stable throughout the loading.}}
    \label{fig:branches-stiff}
\end{figure}

In order to assess the stability of equilibrium branches we numerically evaluate of the smallest eigenvalue of the second variation by discretizing the integrals \eqref{hessian22} and   \eqref{hessian222} to construct the stiffness matrices
$\stiffmat$ and $\widetilde\stiffmat$, investigating numerically the sign of the minimal eigenvalue $\lambda_t$ of the corresponding discrete quadratic form \cite{Sanderson2016-ht}.  The finite element discretization of the displacement and damage field $(u, \alpha )$     with \( n_u \) displacement degrees-of-freedom
$
    \mathbf{u} = \{ u_1, \ldots, u_{n_u} \}^T
$
and \( n_\alpha \) damage degrees-of-freedom
$
    \boldsymbol{\alpha} = \{ \alpha_1, \ldots, \alpha_{n_\alpha} \}^T
$ is given by
$
    u(\mathbf{x}) \approx u_{\text{FE}} (\mathbf{x}) := \sum_{i=1}^{n_u} \mathcal{N}^{(u)}_i(\mathbf{x}) u_i $
and $\alpha(\mathbf{x}) \approx \alpha_{\text{FE}} (\mathbf{x}) := \sum_{i=1}^{n_\alpha} \mathcal{N}^{(\alpha)}_i (\mathbf{x}) \alpha_i
$
where $\mathcal{N}^{(u)}(\mathbf{x}) $ and $\mathcal{N}^{(\alpha)}(\mathbf{x}) $ are the finite element basis functions and {$u_i$} and {$\alpha_i$} are nodal values for the displacement and damage fields, respectively. {Because both damage and displacement fields are discretised using the same finite element basis functions, we henceforth denote them by the same symbol $\mathcal N_i$ for notational convenience.}

    {
        Using a hat $\hat{\mathcal{N}}_j(\xi)$ to denote the basis functions defined on the isoparametric domain $\xi \in [-1, 1]$, where $j$ is the local node index within the reference element, the assembly process through which the local basis functions $\hat{\mathcal{N}}_j(\xi)$ are connected to the global finite element fields is the mapping
        $$u_{\text{FE}}(x) = \sum_{i=1}^{n_u} \mathcal N_i(x) u_i = \sum_{{\mathrm{e}}=1}^{n_\text{ele}} \sum_{j=1}^{3} \hat{\mathcal {N}}_j(\xi) u_j^e,$$
        where $u_j^e$ is the value at the $j$-th node of the restriction of the field $u$ to the element ${\mathrm{e}}$, and $u_{\text{FE}}(x)$ is the field in the global physical domain.}

For quadratic one-dimensional finite elements with 3 nodes, shape functions   ${\mathcal N}_i(x)$ at the nodes  are given by ${\mathcal N}_1(\xi)=-0.5\xi(1-\xi)$, ${\mathcal N}_2(\xi)=-0.5\xi(1+\xi)$ and ${\mathcal N}_3(\xi)=-(1-\xi)(1+\xi)$, where $\xi$ is the isoparameteric coordinate. The discrete solution $u(x_i)$ provided at discrete nodes $x_i$ by AUTO was first interpolated using B-spline basis functions of degree 3 \cite{Grimstad2016-cq}, and then used to calculate the integrals \eqref{hessian22} and \ref{hessian222} employing a three-point Gauss integration scheme. The fixed boundary conditions were enforced by removing the rows and columns corresponding to $x = 0$ and $x = 1$ from the stiffness matrices
$\stiffmat$ and $\widetilde\stiffmat$, {which are real, symmetric, block matrices of dimensions ${n_u + n_\alpha}$ and ${2(n_u) + n_\alpha}$, respectively}. The explicit form of the stiffness matrix for the stiff substrate model is given by
\begin{equation}
    \stiffmat =
    \begin{bmatrix}
        \stiffmat_{uu}       & \stiffmat_{u\alpha}      \\
        \stiffmat_{\alpha u} & \stiffmat_{\alpha\alpha}
    \end{bmatrix}
    =
    \begin{bmatrix}
        \int_0^1[ \frac{1}{\Lambda^2}{\mathcal N}_i{\mathcal N}_j + (1-\alpha)^2{\mathcal N}'_i{\mathcal N}'_j
        ] dx &
        -2\int_0^1(1-\alpha)u' {\mathcal N}'_i {\mathcal N}_{{j'}}  dx                                                            \\
        -2\int_0^1(1-\alpha)u' {\mathcal N}_i {\mathcal N}'_j dx
             & \int_0^1 [ (2+u'^2){\mathcal N}_{{i'}}{\mathcal N}_{{j'}} +\damagell^2{\mathcal N}'_{{i'}}{\mathcal N}'_{{j'}}] dx
    \end{bmatrix},
    \label{eq:stifness1}
\end{equation}
whereas for the compliant substrate model it reads
\begin{multline}
    \stiffmatcompl=
    \begin{bmatrix}
        \stiffmatcompl_{uu}       & \stiffmatcompl_{u\alpha}      & \stiffmatcompl_{uv} \\
        \stiffmatcompl_{\alpha u} & \stiffmatcompl_{\alpha\alpha} & \textbf{0}          \\
        \stiffmatcompl_{vu}       & \textbf{0}                    & \stiffmatcompl_{vv}
    \end{bmatrix}
    \\=
    \begin{bmatrix}
        \int_0^1[ \frac{1}{\Lambda^2}{\mathcal N}_i{\mathcal N}_j + (1-\alpha)^2{\mathcal N}'_i{\mathcal N}'_j] dx         &
        -2\int_0^1(1-\alpha)u' {\mathcal N}'_i {\mathcal N}_{{j'}}  dx                                                     &
        -\int_0^1[\frac{1}{\Lambda^2}{\mathcal N}_i {\mathcal N}_{{k}}]  dx                                                                                                                     \\
        -2\int_0^1(1-\alpha)u' {\mathcal N}_{{i'}} {\mathcal N}'_j dx                                                      &
        \int_0^1 [ (2+u'^2){\mathcal N}_{{i'}}{\mathcal N}_{{j'}} +\damagell^2{\mathcal N}'_{{i'}}{\mathcal N}'_{{j'}}] dx &
        \textbf{0}                                                                                                                                                                              \\
        -\int_0^1[\frac{1}{\Lambda^2}{\mathcal N}_{{l}}{\mathcal N}_j ] dx                                                 & \textbf{0} & \int_0^1 \rho {\mathcal N}_{{k}}{\mathcal N}_{{l}} dx
    \end{bmatrix},
    \label{eq:stifness2}
\end{multline}
{where $i, j = 0,\dots, n_u; i', j' = 0,\dots, n_\alpha, $ and $k, l = 0,\dots, n_v$ are the degrees of freedom indices for the film displacement, damage, and foundation displacement, respectively.}
Our objective is to establish a branch switching strategy that,
as the external loading parameter monotonically increases, allows equilibrium branch transitions when the current branch ceases to be stable or to exist.
This strategy must ensure the system's re-stabilization following an instability in a dissipative manner. In a quasi-static scenario, it should select a new locally stable equilibrium branch with inherently lower energy.  Considering applications in structural mechanics, our approach to selecting the new equilibrium branch relies on a criterion of local energy minimization (LEM) which emulates the zero viscosity limit of overdamped viscous dynamics. This approach differs from a global energy minimizing (GEM) strategy, which may be more relevant in biomechanical applications~\cite{Salman2021-mn}. According to LEM protocol, during quasi-static loading, the system will remain in a metastable state (a local minimum of energy) until it becomes unstable. Subsequently, during an isolated switching event, the new equilibrium branch will be chosen using a descent algorithm \cite{Puglisi2005-lg}.
\begin{figure}
    \centering
    \hspace*{-.3cm}
    \includegraphics[align=c,width=.7\textwidth]{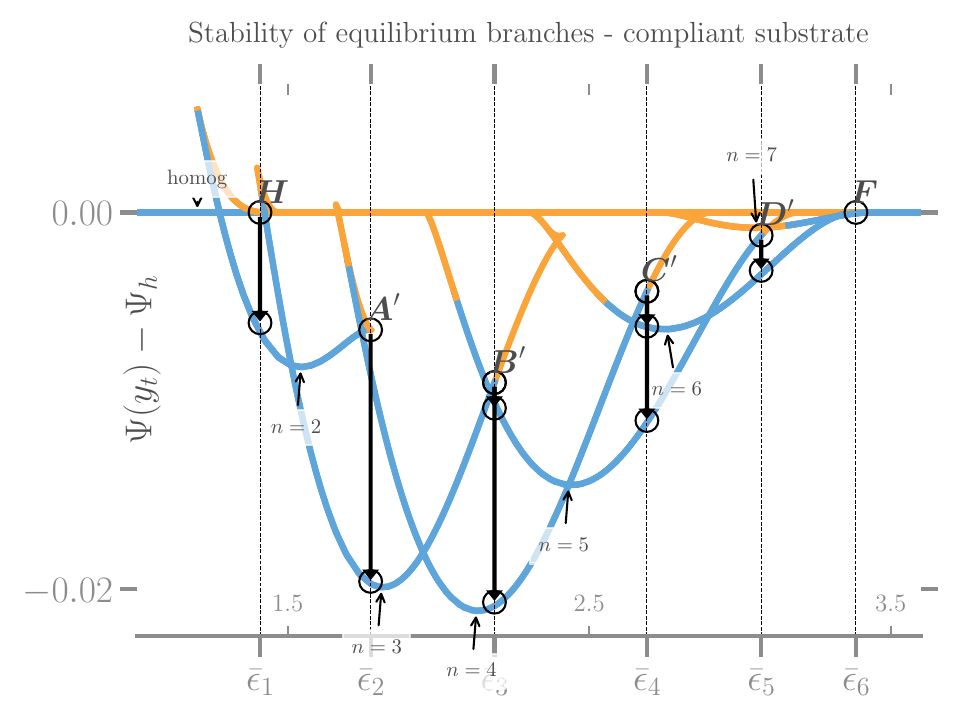}
    \includegraphics[align=c,width=.3\textwidth]{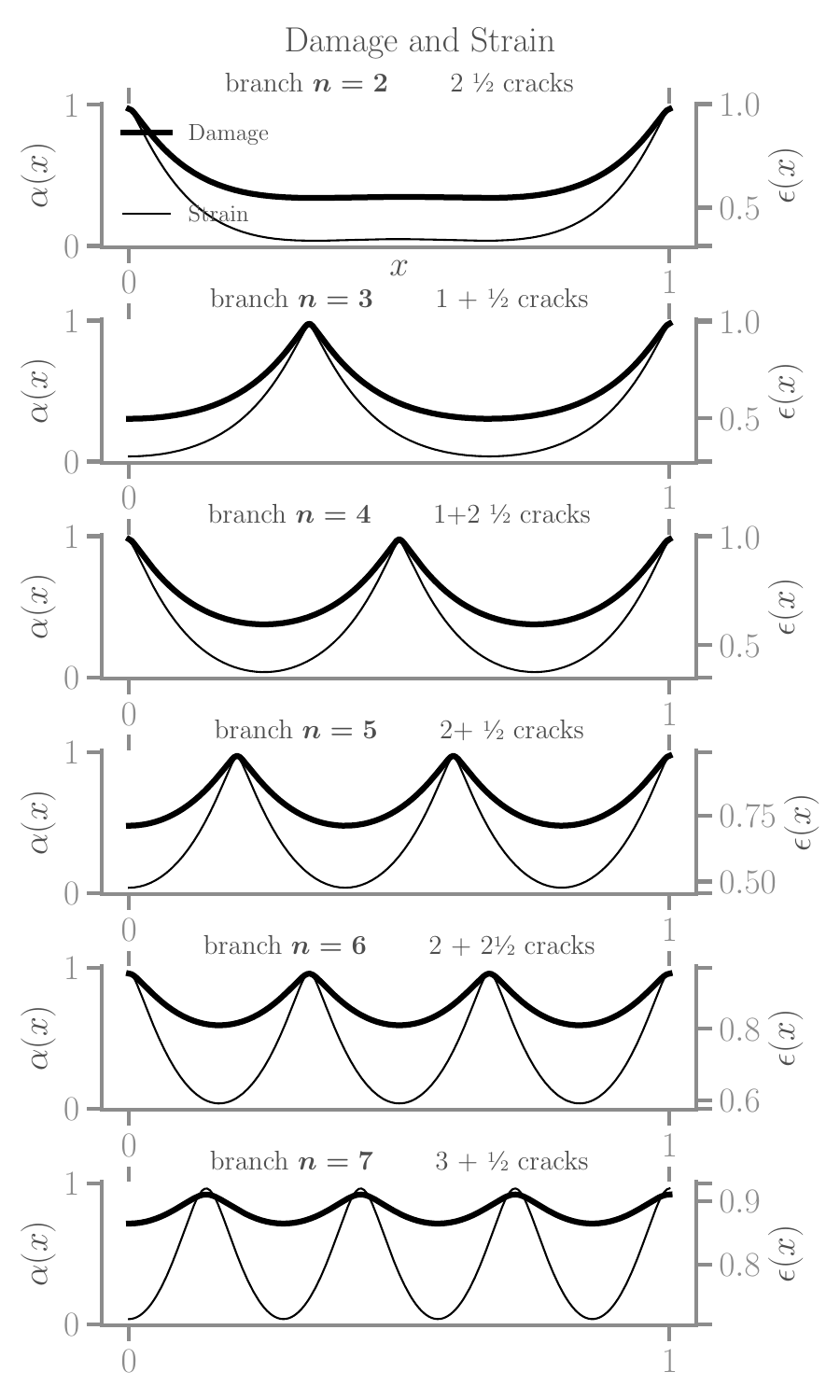}
    \includegraphics[align=c,width=.8\textwidth]{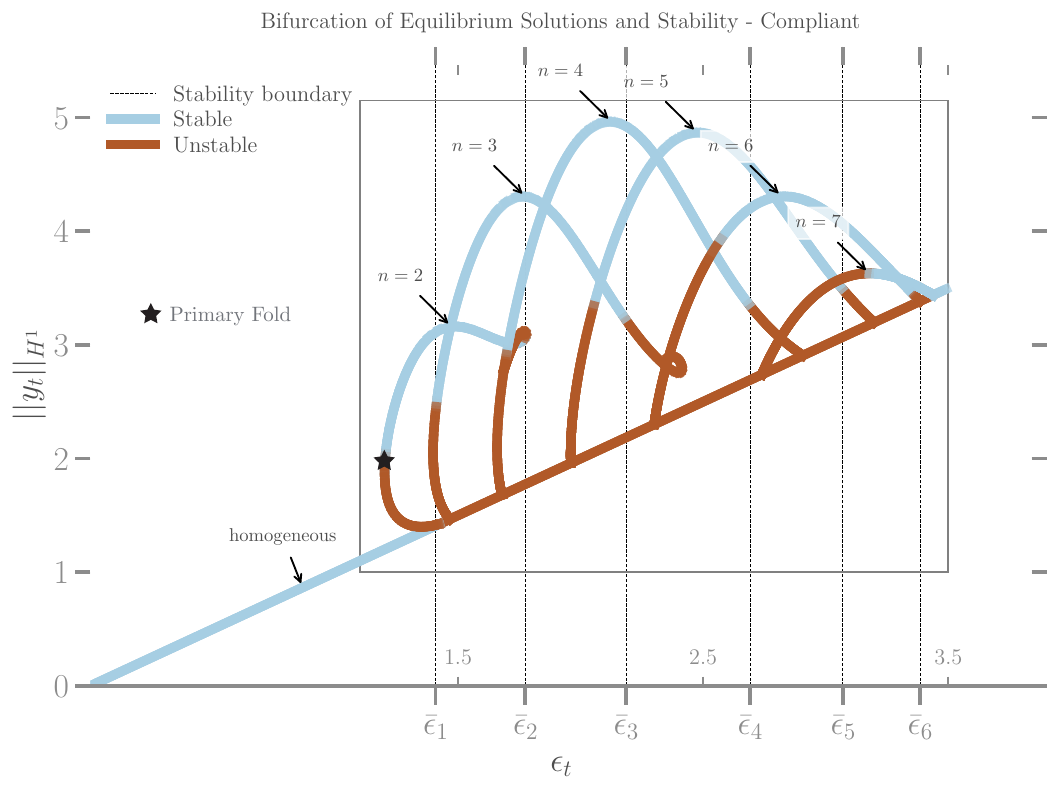}
    \caption{{Equilibrium branches for phase-field thin film model on compliant elastic foundation. Left: energy difference $\Delta\Psi$ between current and homogeneous configurations at equal load. Stability color-coded: blue (stable), orange (unstable) based on smallest eigenvalue of stiffness matrix $H$. Branches parametrised by integer $n$; arrows indicate branch switching at stability loss. Right: damage and strain profiles of minimum energy configurations on each branch. Bottom: bifurcation and stability map. Subcritical bifurcation and primary fold in the presence of a compliant substrate, introducing additional complexity and multiple secondary folds and bifurcations (not investigated). See also the caption of Figure~\ref{fig:branches-stiff}.}}
    \label{fig:branches-compliant}
\end{figure}

Figures~\ref{fig:branches-stiff},~\ref{fig:branches-compliant} show equilibrium branches that solve the nonlinear equations~\ref{auto1} and ~\ref{auto2}. These branches are represented in {the top-left subfigure} by plotting the energy difference $\Delta \Psi$ between the energy of the current solution and the energy of the homogeneous solution $\Estiffhom(\bar{\epsilon}_t)$ at the current value of the loading parameter $\bar\epsilon_t$.
Stability is discerned through numerical evaluation of the smallest eigenvalue $\lambda_t$ of the stiffness matrix $\stiffmat$ (and $\stiffmatcompl$) defined in \eqref{eq:stifness1} (respectively, in \eqref{eq:stifness2}).
{The bifurcation points and associated wavelengths predicted by the linear stability analysis match, up to numerical precision, with those obtained from AUTO analysis for the different values of $n$.}

Because eigenvalues represent the local curvatures of the energy functional, the sign of the smallest  $\lambda_t$ determines the stability of the solution, with {${\inf}\lambda_t > 0$} indicating stability and $\inf\lambda_t < 0$ indicating instability.

Under the LEM protocol, the system explores the fundamental branch which is stable until
$\bar\epsilon_1$, identified using linear stability analysis, see Figure~\ref{fig:branches-stiff}.
At {this} critical load on the homogeneous branch  at point $H$, a first instability determines a branch switching transition, from the {homogeneous} branch with wavenumber $n = 0$ to the nontrivial equilibrium branch with $n = 3$. The ensuing non-homogeneous configuration is linearly stable, as seen in Fig. \ref{fig:branches-stiff}. {The} lowest energy configuration on this equilibrium branch (cf. point $A$ in Fig.~\ref{fig:branches-stiff}) consists of two simultaneously nucleated localized cracks, one inside the domain and one on the boundary.

When the loading parameter $\bar{\epsilon}_t$ is further increased, the equilibrium configuration with $n=3$ loses linear stability at point $A'$. Transitions according to the LEM protocol {show that}
only available transition from point $A'$ is to the branch with $n=4$, which is locally stable within at the corresponding applied strain $\bar{\epsilon}_2$, {see Fig.~\ref{fig:branches-stiff}.}. The minimum energy configuration on the branch with $n=4$ consists of two cracks: one inside the domain and two on the right and left boundaries, see {point $B$} in Fig.~\ref{fig:branches-stiff}-right.
Further increasing the load, the stability of the branch with $n=4$ is lost at point $B'$, a single available transition leads the system to the branch with $n=5$. While the branch with $n=6$ appears accessible due to its lower energy, it is unstable at the current load value $\bar\epsilon_3$. The lowest energy configurations on the branches with $n=5$ and $n=6$ are illustrated in Fig.~\ref{fig:branches-stiff}-(right). Increasing the load along the branch with $n=5$ which loses linear stability at point $C'$, a branch switching event occurs towards the branch with $n=7$, ({see the corresponding} configuration in Fig.~\ref{fig:branches-stiff}-(right)). Finally, this branch reconnects to the homogeneous branch ($n=0$) at point $F$. Remark that, the LEM protocol identifies for the stiff substrate case a \emph{unique evolution path} despite  multiplicity of equilibrium solution, with the a single bifurcation option, or available branch, at each instability point.

For the compliant substrate model, the equilibrium branches that solve the nonlinear equations (Eq.~\ref{auto2}) are shown in Fig.~\ref{fig:branches-compliant}.
Again, we plot the energy difference $\Delta \widetilde\Psi$ between the energy of the current solution and the energy of the homogeneous solution $\Estiffhom (\bar{\epsilon})$ at the current value of the loading parameter $\bar\epsilon$.

According to the LEM protocol with a loading history starting in the unloaded and sound configuration, the initial transition from the trivial solution to the only available branch with $n=2$ occurs at $\bar \epsilon_1$, marked as point $H$.
The lowest energy configuration {after transition} at point $A$ consists of two boundary cracks,
{
        see the the damage and strain profiles illustrated in Fig. \ref{fig:branches-compliant}-(right)
    }. As the loading increases, the system persists on the $n=2$ branch until point $A'$ which marks an instability.
The system can now access two equilibrium branches: $n=3$ and $n=4$.
These feature one bulk crack plus one boundary crack, and one bulk crack plus two boundary cracks, respectively, {see
        Fig.~\ref{fig:branches-compliant}-(right) illustrating the typical damage and strain profiles on these branches.}

The choice of the subsequent branch from point $A'$ will dictate the ensuing crack growth. On branches with $n=3$ and $n=4$, at the instability points $B'$ and $C'$, the system will once again encounter two available equilibrium branches. From $n=3$ branch  at point $B'$, the system can transition to either the $n=4$ or $n=5$ branch. Opting for the $n=4$ branch, the subsequent branch selection occurs at point $C'$, offering branches with $n=6$ or $n=5$. Notably, the {$n=6$} branch smoothly reconnects with the trivial solution at point $F$. However, the {$n=5$} branch experiences another instability at point $D'$ before rejoining the trivial solution.

    {
        Diagrams~in Fig~\ref{fig:branches-stiff}-(bottom) and~in Fig~\ref{fig:branches-compliant}-(bottom) (for the compliant system) offers further qualitative insight of the bifurcation behaviour.
            {The} stiff system (considerations in the compliant case are, at first order, analogous),
        evolves starting from the zero state along the stable homogeneous branch, exhibiting a subcritical pitchfork bifurcation at load $\bar \epsilon_1$ (at point $H$).
        Pitchfork bifurcations occur in problems with symmetry an symmetry breaking, which is the
        case for the homogeneous solution curve.
        The  first critical load {marks a subcritical bifurcation}, evidenced by the fundamental branch becoming unstable beyond $\bar \epsilon_1$, accompanied by the emergence of a bifurcating branch.
        Several additional bifurcations from the homogeneous branch occur at loads $\bar \epsilon_i, i\leq 5$ without change in stability. The homogeneous solution eventually regains stability at the (supercritical) pitchfork point $F$, corresponding to the load $\bar \epsilon_5$.
        The first emergent branch arising a $\bar \epsilon_1$, corresponds to states of higher energy compared to the homogeneous states, confirming its unstable character (see energy diagram in Figure~\ref{fig:branches-stiff}, top-left panel)
        The solution curve folds back at the turning point $\bigstar$, with a restoration of stability.
        This stability switch at the fold is characterized by a vertical tangent to the solution curve, where the slope reverses, and the stable and unstable branches merge. Consequently, the Hessian becomes singular at this point, with exactly one eigenvalue transitioning from negative to positive, indicating a change in stability without the emergence of additional (secondary) branches.
        Further along the bifurcated localized branch $n=2$, two additional bifurcations occur, giving rise to two secondary branches. These secondary bifurcation points are indicated with black borders on the primary bifurcated branch. Notably, the first secondary branch contains a stable segment followed by a corner fold without any accompanying change of stability, while the second secondary branch is entirely unstable.
        Similar to the homogeneous branch, some bifurcations occur without affecting stability, indicating the onset of zero eigenvalues alongside existing negative ones.
        The compliant system (Figure~\ref{fig:branches-compliant}, bottom) exhibits analogous behavior but with increased complexity due to additional degrees of freedom associated with substrate deformation. This complexity results in a richer bifurcation structure, featuring multiple turning points, some accompanied by stability changes and others without.
    }

For the current choice of material parameters, the compliant substrate model shows more branch switching events and {notably,} non-uniqueness of the global response.
    {Unlike} the stiff substrate model {where} the LEM strategy identifies a unique evolution path, the compliant model allows \emph{eight} different trajectories in phase-space, given by the various available path-bifurcation choices.
This additional richness provides a good test case for numerical optimization methods, which we will discuss in the following.

\subsection{Equilibrium branch selection by overdamped viscous dynamics}
We consider now the equilibrium solutions reachable through the line-search based quasi-Newton algorithms such as  conjugate gradient or the BFGS optimization, which effectively  mimic the zero viscosity limit of overdamped viscous dynamics~\cite{SALMAN2012219}. Quasi-Newton methods serve as alternatives to Newton's method for locating roots or local extrema of functions. Particularly advantageous in scenarios where computing the Hessian at each iteration is impractical or computationally expensive, these methods circumvent the need for explicit computation of energy derivatives. Instead, they rely on evaluating the function value and its gradient and updating the Hessian by analyzing successive gradient vectors.

Quasi-Newton methods are highly suitable for solving the phase-field equation of fracture, particularly when compared to standard Newton method-based monolithic solvers. Such solvers, which simultaneously  solve the equations for both damage and displacement variables, often falter when confronted with nonconvex energy functionals. For example, as demonstrated in \cite{Wick2017-bo}, the Newton method-based monolithic algorithm does not consistently handle brittle fracture scenarios involving abrupt crack propagation. Recently, quasi-Newton methods, particularly the BFGS variant, have been employed to effectively solve the system of coupled governing equations in a monolithic fashion within the phase-field method of fracture. These methods have demonstrated success in various engineering applications, as evidenced by \cite{Kristensen2020-zy,Wu2020-qk,Salman2021-mn,Liu2022-ix}.

Our primary objective is not to provide a comprehensive assessment of quasi-Newton methods on a global scale. Instead, our focus lies in scrutinizing their behavior and performance specifically concerning equilibrium branch selections within our simplified framework, where all branches are readily identified. By narrowing our scope to this specific aspect, we aim to gain insights into the effectiveness and reliability of quasi-Newton methods in reaching stable states of our system.

\begin{figure}
    \centering
    \hspace*{-.3cm}
    \includegraphics[width=.6\textwidth]{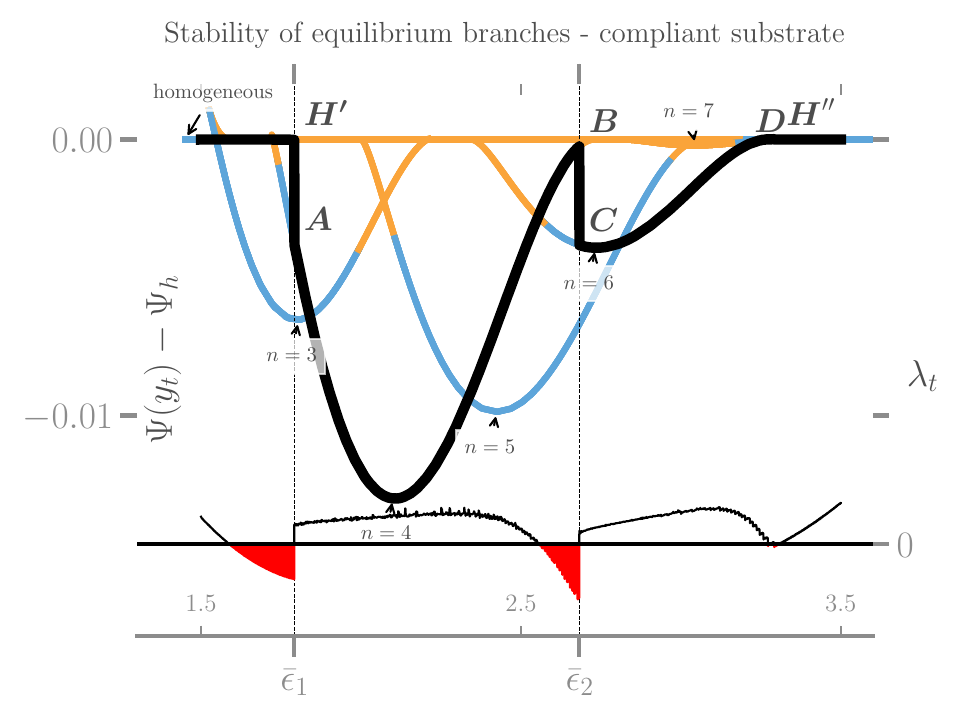}
    \includegraphics[width=.4\textwidth]{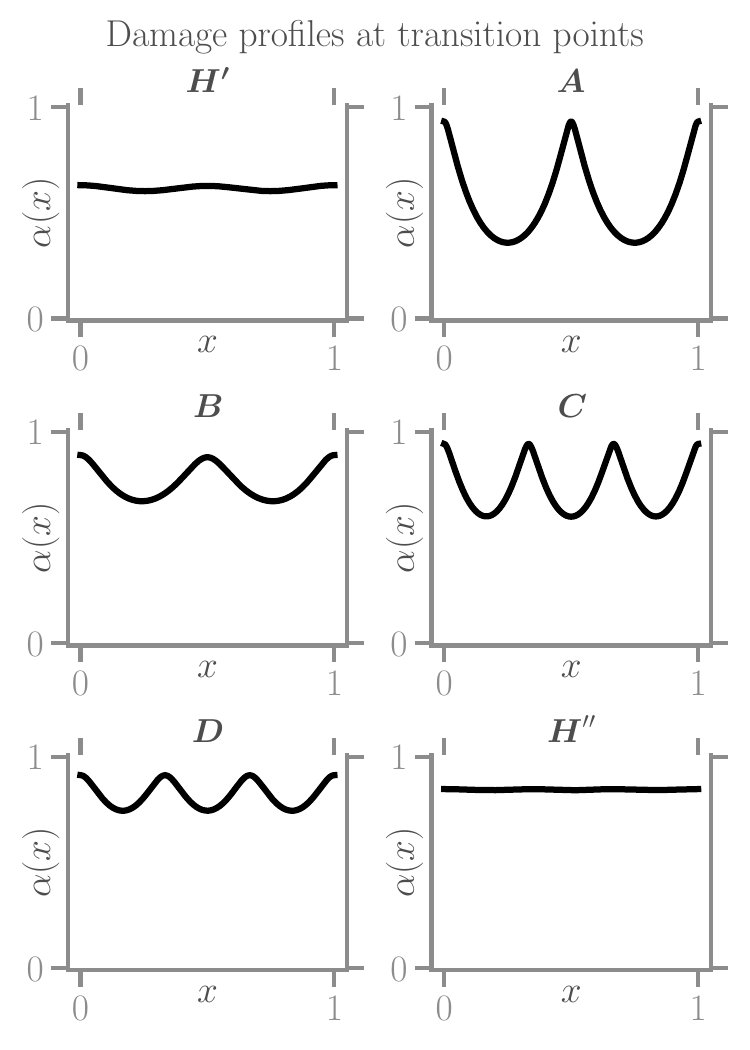}
    \caption{{Stiff substrate. Quasi-static L-BFGS simulations: (a) energy difference $\Delta\Psi$ between quasi-Newton and homogeneous solutions superimposed on equilibrium branches (light blue: stable; orange: unstable). Bottom: smallest eigenvalue $\lambda_t$ of second variation $\Psi''$ vs.\ loading parameter $\bar{\epsilon}$. Red indicates unstable region ($\lambda_t < 0$) where solver solutions violate evolution law. Right: damage profiles at transition end states.}
    }
    \label{fig:tempo1}
\end{figure}
Recall that quasi-Newton algorithms only evaluate the function value and its gradient to reach equilibrium configurations. This implies that in our case, we need to evaluate integrals \eqref{def:energy_stiff} and \eqref{def:energy_compliant}, along with their first variations given by \eqref{firstvar1} and \eqref{firstvar}, in the stiff and compliant models, respectively. We   discretised the integrals \eqref{firstvar1} and \eqref{firstvar} to construct the residuals vectors using finite elements to obtain
    {
        \begin{equation}
            \mathbf{R} =
            \begin{bmatrix}
                \mathbf{R}_u \\
                \mathbf{R}_\alpha
            \end{bmatrix}
            =
            \begin{bmatrix}
                \int_0^1 \left[
                    {{\soften}(\alpha)} u' \mathcal{N}_i' + \frac{1}{\Lambda^2} (u - \tilde{u}) \mathcal{N}_i
                \right] dx \\
                \int_0^1 \left[
                (\frac{1}{2} u'^2 {\soften}'(\alpha) + \homogdiss'(\alpha)) \mathcal{N}_{j'} + \ell^2 \alpha' \mathcal{N}'_{j'}
                \right] dx
            \end{bmatrix}.\label{residual1}
        \end{equation}
    }
where $i = 0, \dots, n_u$, and $j' = 0, \dots, n_\alpha$
in the stiff model and

    {
        \begin{equation}
            \widetilde{\mathbf{R}} =
            \begin{bmatrix}
                \mathbf{R}_u      \\
                \mathbf{R}_\alpha \\
                \mathbf{R}_{{\subsu}}
            \end{bmatrix}
            =
            \begin{bmatrix}
                \int_0^1 \left[
                    {{\soften}(\alpha)}u' \mathcal{N}_i' + \frac{1}{\Lambda^2} (u - {\subsu}) \mathcal{N}_i
                \right] dx \\
                \int_0^1 \left[
                (\frac{1}{2} u'^2 {\soften}'(\alpha) + \homogdiss'(\alpha)) \mathcal{N}_{j'} + \ell^2 \alpha' \mathcal{N}'_{j'}
                \right] dx \\
                \int_0^1 \left[
                    \rho {\subsu}' \mathcal{N}'_i + \frac{1}{\Lambda^2} (u - {\subsu}) \mathcal{N}_i
                    \right] dx
            \end{bmatrix}.
            \label{residual2}
        \end{equation}
    }
where $i, k= 0, \dots, n_u$, and $j' = 0,\dots, n_\alpha$
in the compliant substrate model.

Among iterative methods for large-scale unconstrained optimization, particularly when dealing with possibly dense Hessian matrices,  quasi-Newton methods often emerge as preferable alternatives to the widely-used Newton-Raphson (NR) algorithm. The latter, conventionally utilised for solving linear equations to determine the correction $\Delta \mathbf{X}^{(k)}$ from the current estimate $\mathbf{X}^{(k)} = (\mathbf{u}^{(k)}, \boldsymbol{\alpha}^{(k)})$ at iteration $k$, is expressed in our context as:
\begin{equation}
    {H}_{ij} \Delta X_j^{(k)} + R_i = 0,
    \label{Eq:NR}
\end{equation}
where the discrete stiffness matrix $\mathbf{H}$ and bulk forces $\mathbf{R}$ are {bootstrapped} with the initial guess $\mathbf{X}^{(k)}$. Subsequently, the guess is updated as $\mathbf{X}^{(k+1)} = \mathbf{X}^{(k)} + \Delta \mathbf{X}^{(k)}$ after solving Equation \eqref{Eq:NR} using LU factorization~\cite{Sanderson2016-ht}. Cleary, the NR algorithm fails if the discrete stiffness matrix $\mathbf{H}$ is not invertible.

On the other hand, quasi-Newton methods are well-established (see standard textbooks, e.g., \cite{Nocedal1999-zr,Nocedal2006-qh}), and generate a sequence $\left\{\mathbf{X}^{(k)}\right\}$ according to the following scheme:
\begin{equation}
    \mathbf{X}^{(k+1)} = \mathbf{X}^{(k)} + h^{(k)} \mathbf{p} ^{(k)}, \quad k=0,1,\ldots
\end{equation}
with
\begin{equation}
    \mathbf{p}^{(k)}=-(\mathbf{B}^{(k)})^{-1}{\bf R},
\end{equation}
where $(\mathbf{B}^{(k)})^{-1}$ approximates the inverse of the Hessian matrix  $\mathbf{H}$ and $h^{(k)}$ represents a step length. Particularly, instead of computing $(\mathbf{B}^{(k)})^{-1}$  at each iteration $k$, these methods update $(\mathbf{B}^{(k)})^{-1}$ in a straightforward manner to obtain the new approximation $(\mathbf{B}^{(k+1)})^{-1}$  for the subsequent iteration. Additionally, rather than storing full dense approximations, they only retain a few vectors of length $N=n_\alpha + n_u$, enabling implicit representation of the approximations. The choice of the step length $h^{(k)}$ is carried out through a line search to minimise a function $f(h) = f(\mathbf{X}^{(k)} + h \mathbf{p}^{(k)})$ in order to find an acceptable step size $h^{(k)}$ such that $h^{(k)} \in \arg \min_{h} f$.

Among quasi-Newton schemes, the L-BFGS method is widely regarded as one of the most efficient and well-suited for large-scale problems due to its limited and user-controlled storage requirements. This method constructs an approximation of the inverse Hessian matrix, leveraging curvature information solely from recent iterations. Note also that the update formula for the approximative in successive minimization steps depends on the adapted algorithm. These algorithms are extensively used in the literature, and we refer the reader to the references for a more detailed description~\cite{Matthies1979-gl,Xu2001-ax,Nocedal1999-zr,Nocedal2006-qh,Simone2012-tx,Lewis2013-eu,Curtis2015-wp}. However, quasi-Newton methods do present certain drawbacks, notably slow convergence for ill-conditioned problems, particularly when the eigenvalues of the Hessian matrix are widely dispersed~\cite{Simone2012-tx}.
\begin{figure}
    \centering
    \hspace*{-.3cm}
    \includegraphics[width=.6\textwidth]{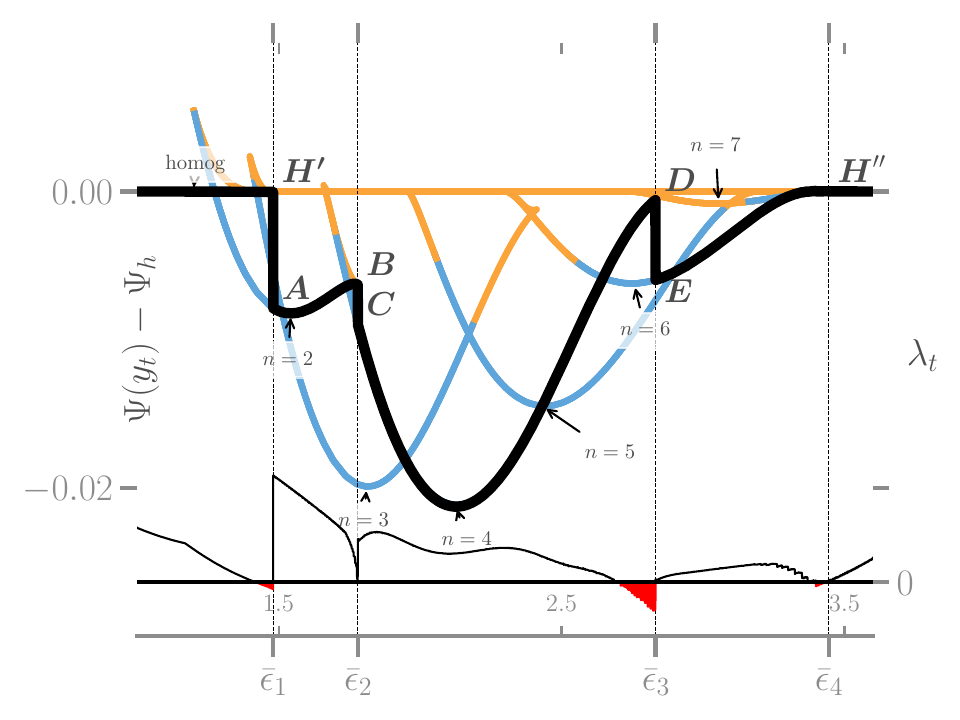}
    \includegraphics[width=.4\textwidth]{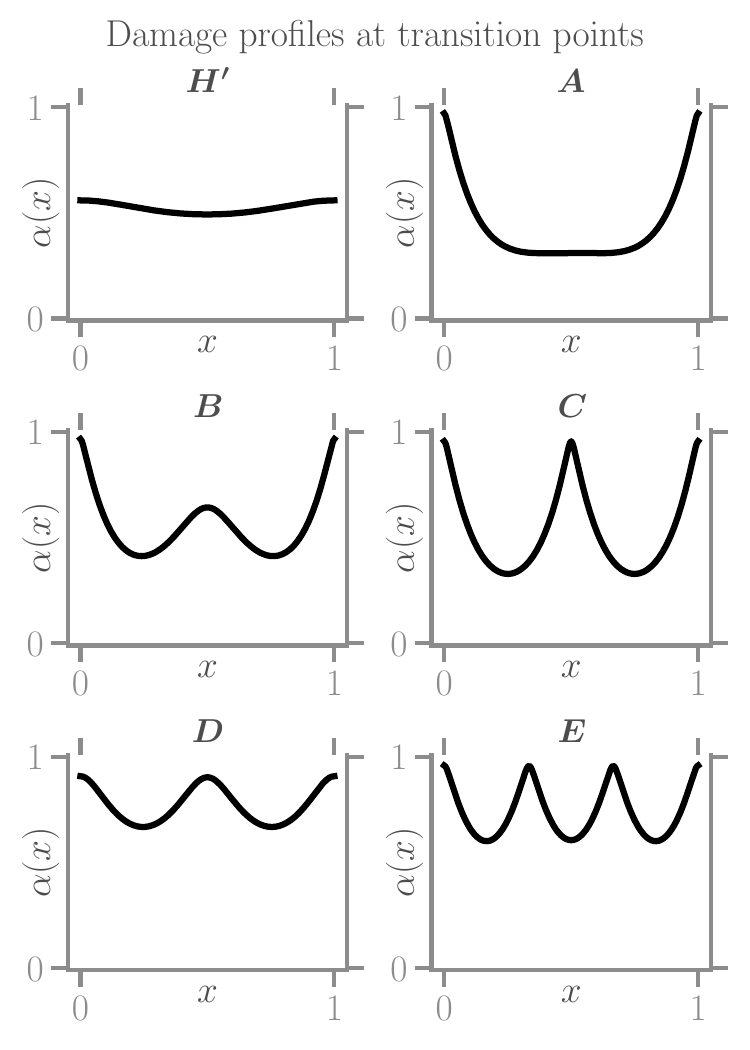}
    \caption{{Compliant substrate. Quasi-static L-BFGS simulations: (a) energy difference $\Delta\Psi$ between quasi-Newton and homogeneous solutions superimposed on equilibrium branches (light blue: stable; orange: unstable). Bottom: smallest eigenvalue $\lambda_t$ of second variation $\Psi''$ vs.\ loading parameter $\bar{\epsilon}$. Red indicates unstable region ($\lambda_t < 0$) for equilibrium solution. Right: damage profiles at transition endstates.}}
    \label{fig:tempo2}
\end{figure}

In our numerical experiments, we employ the BFGS and L-BFGS solvers from the Python SciPy library~\cite{2020SciPy-NMeth} and the Alglib library~\cite{Bochkanov2013-lk}. These solvers utilise the residual vectors (see \ref{residual1} and \ref{residual2}) at each finite element node, alongside the values of integrals \eqref{def:energy_stiff} and \eqref{def:energy_compliant}.
The following results were obtained using the Alglib library~\cite{Bochkanov2013-lk}; however, our results are consistent across other numerical implementation of quasi-Newton methods in the python {SciPy} library.

\paragraph{The stiff system.}
In our numerical experiments for the stiff system, we overlay solutions obtained with the quasi-Newton algorithm onto the equilibrium map determined using the pseudo-arclength continuation method described in the previous section, {see Figure~\ref{fig:tempo1}}. The smallest eigenvalue $\lambda_t$ whose positivity informs on the stability of the solution {is shown in Fig. \ref{fig:tempo1}-bottom}.
All the branch switching events are delayed in the sense they do not take place when the smallest eigenvalue of the second variation vanishes but only beyond the loads corresponding to the eigenvalue's sign transition.
Notably, the expected first branch switching event --- from the {homogeneous} solution to the branch with \(n=3\) as outlined in the LEM protocol --- did not occur at point $H$ as anticipated.
During a monotonic loading, the system remains on the trivial branch, as shown in Fig. \ref{fig:branches-stiff}(a), until it transitions to the branch with \(n=4\) at a higher than the critical load \(\bar{\epsilon}_1\).
By monitoring the smallest eigenvalue of the second energy variation at the solution, we observe that indeed the solutions computed via the quasi-Newton solver are unstable beyond \(\bar{\epsilon}_1\), which is consistent with the linear stability results for the trivial branch.
A closer inspection of the solution field reveals that the quasi-Newton solutions are not homogeneous but exhibit a small perturbation akin to the instability mode calculated analytically (i.e., the eigenvector associated with the smallest eigenvalue).
For instance, the  quasi-Newton solution before the first branch switching event shows a slight oscillation reminiscent of the eigenmode \(\alpha_n(x) \sim \cos(n\pi x)\), with \(n=3\), {see the inset (state {$H'$}) in Fig. \ref{fig:tempo1}-right}.
The state transition captured by the quasi-Newton method leads the system onto the \(n=4\) branch, with one bulk crack and two boundary cracks, see inset ($A$) in Fig. \ref{fig:tempo1}-right.
Had the transition taken place at {the loss of stability of the homogeneous solution} as anticipated by our LEM strategy, the system would have reached a configuration with {one} bulk crack and one boundary crack, as shown in Fig. \ref{fig:branches-stiff}-right (branch {$n=3$}).

With continued loading, the system evolves along the \(n=4\) branch beyond the loading value at which stability is lost, as shown in {the stability marker in} Fig. \ref{fig:tempo1}-{left-bottom}.
A  branch switching event takes place from the current branch (\(n=4\)) at point {$B$} to the branch with \(n=6\), see Fig. \ref{fig:tempo1}-(right) and the inset $C$ for the corresponding damage profile.
The delay in bifurcation results in a branch selection event different from the one in our LEM protocol.
The delays at bifurcation for the first and second branch switching events lead to a completely different evolutionary path compared to the LEM protocol until it finally reconnects with the homogeneous branch at point {$H''$}.

\paragraph{The compliant system.}
Solutions of our numerical experiments for the compliant model, obtained from the quasi-Newton algorithm are overlaid onto the equilibrium map {see Fig. \ref{fig:tempo2}-(left)}, where their stability is also shown (see Fig. \ref{fig:tempo2}-bottom diagram).
We observe a less pronounced delay for the bifurcation from the {homogeneous} branch, compared to the stiff substrate model. This occurrs at a value higher than the analytically determined critical load \(\bar{\epsilon}_1\).
Interestingly, the quasi-Newton method switches to the branch with $n=2$, which has lower energy of the branch with $n=3$.
This occurs despite the branch $n=3$ {is closer in energy norm and is} accessible at the current value $\bar\epsilon$. {This is consistent with} the LEM protocol whose first branch selection event lead to the branch with $n=2$ with two boundary cracks.
Even before the first branch-changing event, the quasi-Newton solution {deviates from the homogeneous branch, and the damage profile has} is perturbed by an oscillatory term of the form \(\cos(n\pi x)\) shown in Figure~\ref{fig:tempo2}{(right)-H'}, reminiscent of the eigenmode $\alpha_n$  with \(n=2\).
These perturbations although negligible from the energetic standpoint (see {the superposition between the energy of the computed evolution and the exact total energy of the homogeneous solution in} Figure~\ref{fig:tempo2}{(left)}), {influence branch selection}.

The system moves along the branch with $n=2$  as the load increases, until the stability transition at load $\bar \epsilon_2$, indicated by point {$B$}. Energy minimization brings the system to the branch $n=4$ although the branch with $n=3$ is also accessible and has lower energy.
Figure~\ref{fig:tempo2}-(right) displays the corresponding damage profiles before and after {the branch} switching events  at points {$B$} and $C$.
The final {transition} before the reconnection to the homogeneous branch takes place at point $D$, which is also an unstable configuration.
This transition brings the system to the branch $n=6$ as seen in Fig.~\ref{fig:tempo2}-(right), the corresponding stable damage profile at point $E$ is shown in Fig. \ref{fig:tempo2}-(right).

In summary, while the results of the quasi-Newton minimization simulations exhibit overall similarities in both models, the trajectory taken by the system in the compliant model more closely adheres with the LEM protocol outlined above. This is attributed to the specific energy landscape of the compliant model. In case of indeterminacy of the trajectory, such as when multiple stable solutions exist at a loss of stability, how to identify an evolution direction?

\subsection{Hybrid \emph{kick} algorithm  for branch switching}
\albdelete{Our investigation shows that the s}Solutions generated by the quasi-Newton algorithms deviate from the LEM protocol outlined in the preceding sections. We observe delayed bifurcations and the persistence of unstable solutions, {both affecting} branch selection events along the evolution.  This is indeed related to the fact that the energy landscape is already flat when the determinant Hessian gets close to zero.
Testing the energy expansion~\eqref{eqn:energy-expansion}
along a the direction of a bifurcation mode, that is setting $y-y_t=h p_n$, where $p_n:=(w_n, \beta_n)(x)$ is the $n$-th eigenmode associated with  the Hessian and $0<h\in \mathbb R$, we obtain that, for admissible states $y$ $\delta$-close to the equilibrium $y_t$ we can write the following lower bound
$$
    \Psi(y)-\Psi(y_t)\geq \frac{h^2}{2}\lambda_t \|p_0\|^2, \quad \forall y: \|y - y_t\| \leq \delta,
$$
where $\lambda_t$ indicates the smallest eigenvalue and $w_0$ the associated eigenmode. The first order term $\delta\Psi(y_t)(w_n, \beta_n)$ vanishes identically owing to the fact that bifurcation modes are admissible fields for the equilibrium condition.
When the smallest eigenvalue continuously approaches zero, the energy landscape morphs from being locally flat (at first order) and locally convex in all admissible directions (including the directions associated with  the eigenmodes), to loosing local convexity in the one non-trivial direction associated with  the eigenvalue that has changed sign.
In the numerical practice,
the convergence of quasi-Newton algorithms depends on the construction of an (approximate) strictly positive Hessian matrix.
As a consequence, such approximation systematically rules out the ability to capture the change of sign (of the smallest eigenvalue) of the Hessian (its singularity), thus the onset of instability and the instability mode.
This can justify the systematic (and algorithm-dependent) delay of the bifurcation events via the quasi-Newton solver, observed in Fig.~\ref{fig:tempo1} and~\ref{fig:tempo2}.

To address this challenge, we introduce a hybrid approach {that explicitly takes into account the singular mode of the Hessian}. In this method, we continuously monitor the smallest eigenvalue of the complete {Hessian} matrix for the equilibrium solutions obtained from the quasi-Newton algorithm.

\begin{figure}
    \hspace*{-.3cm}
    \includegraphics[width=.5\textwidth]{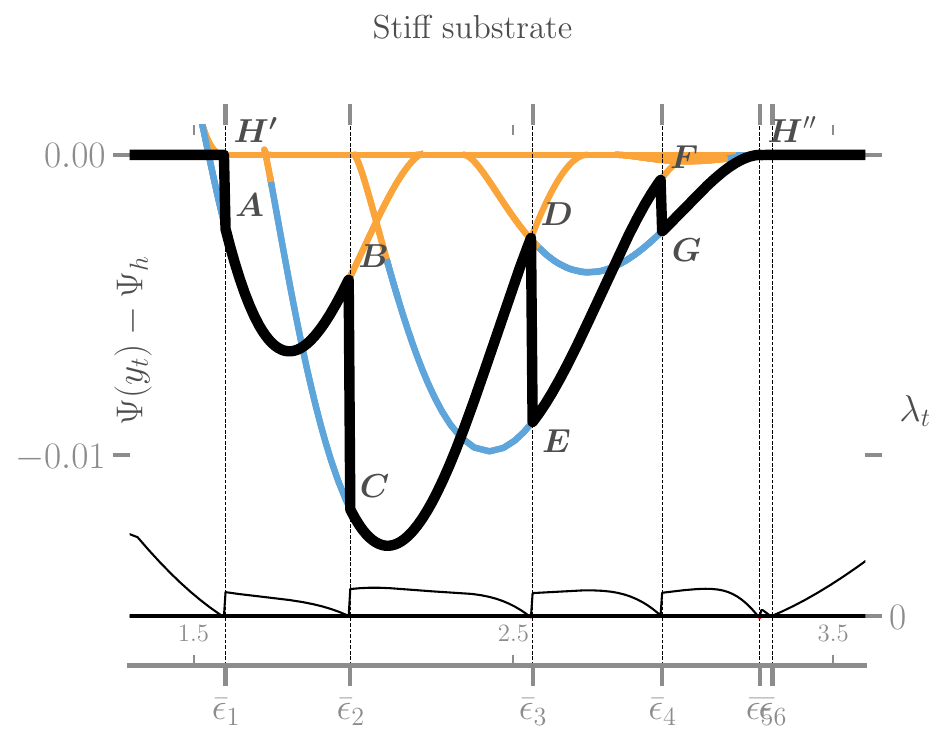}
    \includegraphics[width=.5\textwidth]{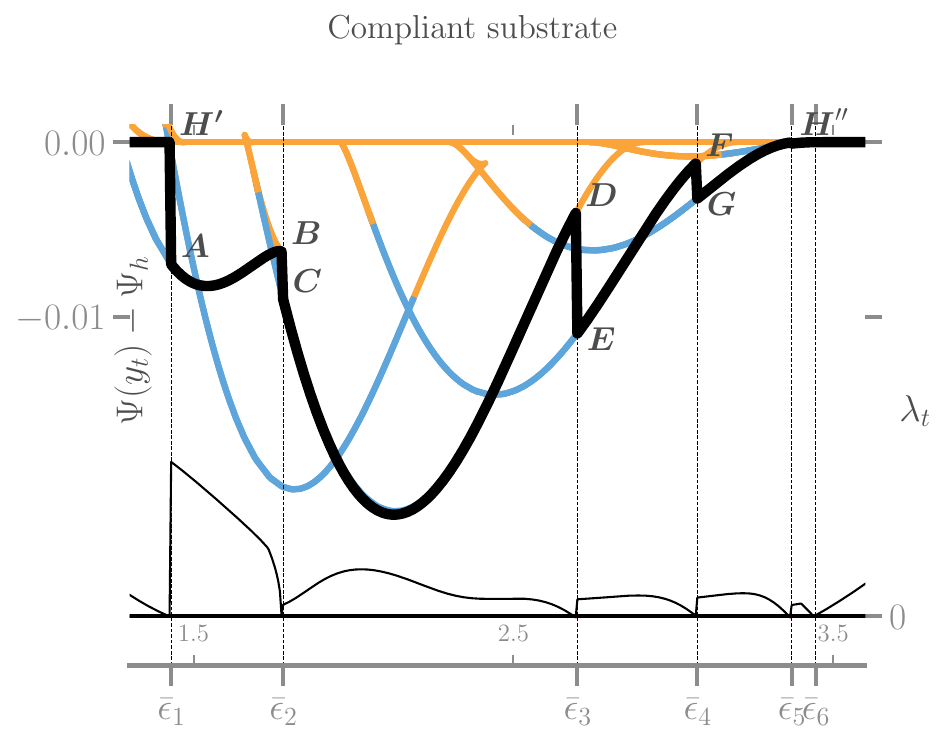}
    \caption{{Quasi-static L-BFGS simulations {with kick algorithm when instability is triggered}: energy difference $\Delta\Psi$ between quasi-Newton and homogeneous solutions superimposed on equilibrium branches (left, {stiff}, and right, {compliant}, models). Bottom: smallest eigenvalue of second variation vs.\ loading parameter $\bar{\epsilon}$. Positive eigenvalues indicate solution and overall evolution stability.}}
    \label{fig:tempostable}
\end{figure}

\begin{figure}[htbp]
    \centering
    \includegraphics*[align=c,width=.45\textwidth,valign=t]{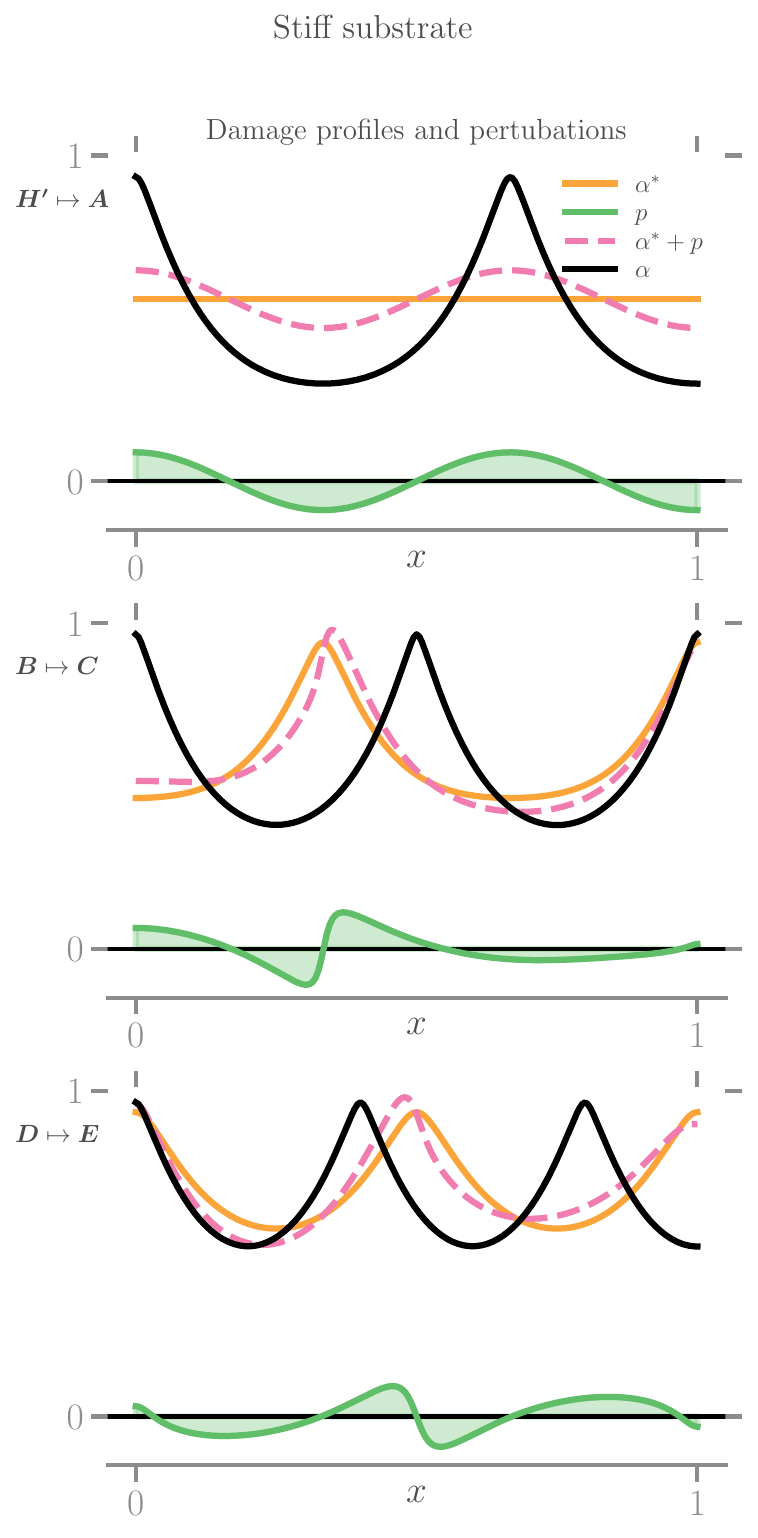}
    \includegraphics*[align=c,width=.45\textwidth,valign=t]{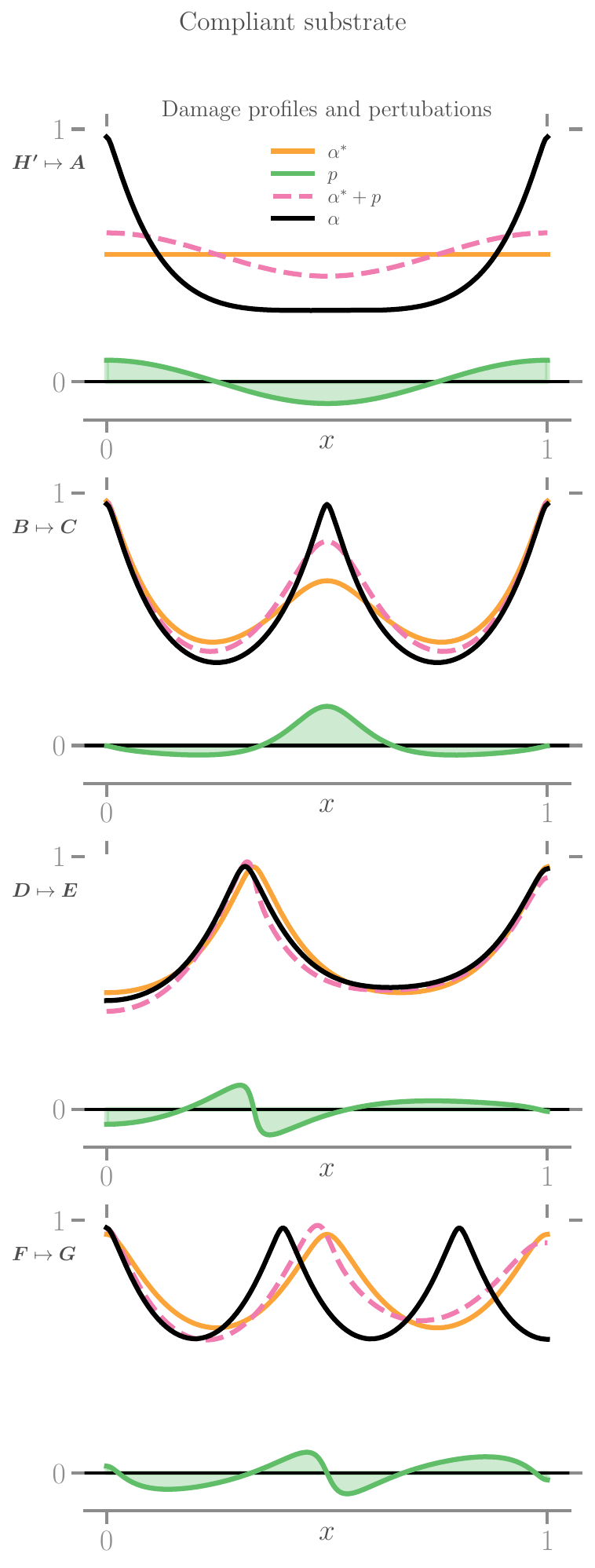}
    \caption{{Profiles of damage fields and perturbations showing kick algorithm effect in state transitions (unstable to stable). Branch switching events correspond to load values $\bar{\epsilon}_i$ with $i \in \mathbb{N}$ in Figs.~7-8. Left column: stiff substrate model; right column: compliant substrate model. Orange line: unstable damage field $\alpha^*$; green line: eigenvector $p$ associated with smallest eigenvalue (positive/negative values highlighted). Dashed line: perturbed damage field $\alpha^* + p$ used as initial guess; solid black line: solution $\alpha_t$ after first-order convergence.}}
    \label{fig:kick}
\end{figure}

When this eigenvalue significantly diminishes, indicating potential instability, we perform a full Newton-Raphson refinement using the solution returned from the quasi-Newton algorithm as an initial guess and obtain the fully homogeneous solution.
Then we calculate the corresponding eigenvector \(\femperturb\), normalised such that $\|\textbf{p}\|=1$.
We then use this eigenvector to perturb the current solution $\femcurrentstate$. This perturbation sets the initial guess for the next minimization step in the quasi-Newton algorithm as $\mathbf{\tilde X}^{(0)} = \femcurrentstate + \eta \femperturb,
$ where \(\eta\) is the step size.
This step size can be determined through a line-search algorithm by minimizing the one-dimensional energy slices along the energy descent mode given by the function
\begin{equation}
    f(\eta) = \Psi(\femstate + \eta \femperturb),
    \label{eqn:energy-slice}
\end{equation}
to find an optimal value \(\eta^*\), such that $\eta^* \in \arg \min_{\eta} f$. {We then run the quasi-Newton step to obtain a new critical point $\femnewstate$.}
    {We now present the computed evolution paths for both stiff and compliant substrates. These trajectories (the thick black lines in Figure~\ref{fig:tempostable}) are globally stable, as confirmed by the positivity of the smallest eigenvalue of the second variation shown in the bottom panels.}
For each {branch-switching event triggered by instability, we display in Fig.~\ref{fig:kick}} snapshots of the {unstable} damage field {$\femcurrentstate$}, the associated {instabilty mode} \(\mathbf{p}\), the perturbed state \(\mathbf{\tilde X}^{(0)}\), and the {post-transition} converged solutions $\femnewstate$, {for the stiff and compliant systems in the left and right columns, respectively}.
\paragraph{The stiff system.}
{T}he {first instability mode} which corresponds to the bifurcation from the homogeneous branch has the form \(\cos(n\pi x)\) with \(n=3\), as identified through the linear stability analysis (see Fig. {\ref{fig:branches-stiff}-right}).
The perturbed quasi-Newton minimization returns a state with two boundary cracks and one interior crack, as shown in Fig. {\ref{fig:kick} {($H'\mapsto A$)}}, which corresponds to the solution found using the arc-length continuation algorithm on the branch with \(n=3\) (see Fig. \ref{fig:branches-stiff}-right).  The hybrid algorithm is also successful in capturing the next two anticipated branch-switching events: from the  branch with \(n=3\)  to the branch  with \(n=4\) {(see Fig. \ref{fig:kick} {($B \mapsto C$)})} and  from branch with \(n=4\)  to the branch  with \(n=5\) {(see Fig. \ref{fig:kick} {($D\mapsto E$)})}. The hybrid algorithm is able to capture the unique path dictated by the LEM protocol shown in Fig. \ref{fig:branches-stiff}.
{The evolution of the smallest eigenvalue during the loading process shows that the solution path is globally stable, see the Hessian inf-spectrum in Fig.~\ref{fig:tempostable}-bottom-left.}
\paragraph{The compliant system.}
{For the compliant system,} the path through which the system may evolve under quasi-static loading is not unique due to the multiplicity of (stable) solutions at critical loads.
Despite the quasi-Newton algorithm being able to reproduce one of the possible paths predicted by the LEM protocol, {the branch switching events} {were delayed with respect to the} anticipated bifurcations.

A stable state with two nucleated boundary cracks (branch \( n=2 \)) follows the destabilization of the homogeneous branch, as illustrated in Fig. \ref{fig:kick} {($H'\mapsto A$)}. Subsequently, an interior crack nucleates in the middle of the system (branch \( n=4 \)), as shown in Fig. \ref{fig:kick} ($B\mapsto C$).
The third event involves a transition from branch \( n=3 \) to branch \( n=5 \). The  event depicted in Fig. \ref{fig:kick} {($D \mapsto E$)} is the transition to branch \( n=6 \). The last event before reconnecting to the homogeneous branch is the transition from branch \( n=6 \) to branch \( n=7 \), the only  branch available to the system at this load, see in Fig. \ref{fig:kick} {($F \mapsto G$)}. To summarise, the branches selected by the hybrid algorithm follow the sequence
\( 0 \rightarrow 2 \rightarrow 3 \rightarrow 5 \rightarrow 6 \rightarrow 0 \),
which differs from the sequence
\( 0 \rightarrow 2 \rightarrow 3 \rightarrow 6 \rightarrow 0 \)
followed by the quasi-Newton algorithm.

Our numerical findings suggest that equilibria close to the Hessian degeneracy (i.e. with small but positive eigenvalues) are robust in the sense that they are insensitive to perturbations. This is highlighed by the fact that running the quasi-Newton solver on a perturbed state before the Hessian sign transition yields, after convergence, the same equilibrium state prior to perturbation.
On the other hand, {at transition points} when the smallest eigenvalue of the system's Hessian is small and negative, the step size \( \eta \) obtained by minimization of \eqref{eqn:energy-slice} is sufficient to effectively escape the flat region of the energy landscape.

\subsection{Irreversible evolutions and stability}
{The algorithm proposed above section is in a branch switching strategy
    similar to that which has been proposed by~\cite{Baldelli2021-gc} by exploiting the concept of the tangent stiffness matrix, obtained by projection of the Hessian on the subset of (linearised) active constraints.
    The main difference with respect to the approach proposed here is that
    we now account for the fully nonlinear stability at second order, in presence of (weakly active) irreversibility constraints, explicitly representing the nonlinearity of the admissible perturbation space.
}
To highlight the role of constraints on the determination of the evolution and its consequences on the space of perturbations, we discuss the full stability problem inequality~\eqref{eq:variational_stability} in our simplified one-dimensional setting through an illustrative numerical computation with the same material parameters as above, enforcing irreversibility.
Irreversibility manifests both as a pointwise constraint and as a global nonlinearity in the perturbation space.
Firstly, it rules out transitions between branches that require a local decrease of damage in favour of a global decrease of energy (e.g. from branch $n=3$ to branch $n=4$, see profiles in the Fig.~\ref{fig:kick}-(left) of the transition {$B\mapsto C$}).
On the other hand, irreversibility is a global constraint that changes the structure of the perturbation space, by only allowing positive perturbations in the damage field.
Consequently, irreversible evolutions are qualitatively different from the unconstrained case.

The following example illustrates the subtle scenario in which two stable irreversible solutions are computed, one of which branches off the homogeneous solution at a bifurcation point without any instability.
Despite lack of uniqueness of the response enabled by the existence of bifurcations, the second computation shows that the system can robustly navigate the purely homogeneous branch, satisfying a sufficient condition for stability (hence for observability) of the computed trajectory.
This juxtaposition underscores a critical sensitivity in that the numerical bifurcation induced by the solver may not reflect an actual instability transition of the physical system, but rather an artifact of the computational method.

    {
        We compute an irreversible evolution using the same parameters as in the rigid substrate model {cf. Figure~\ref{fig:hessian1}(a)},
        using two different numerical methods: first, the quasi-Newton L-BFGS/CG method (via ALGLIB) and second, a Newton solver based on PETSc's reduced-space (active set) algorithms for variational inequalities (via SNES), \cite{petsc-user-ref, petsc-efficient}.
        In the first case, the numerical solution of first order optimality conditions with irreversibility constraints is tackled leveraging the \texttt{minbleic} subpackage of the Alglib library~\cite{Bochkanov2013-lk}, via an active set method that handles inequality constraints in~\eqref{eq:eq_variational_inequality_full}.
        Active inequality constraints correspond to nodes where damage necessarily evolves. Indeed, owing to the complementarity conditions~\eqref{eq:complementarity}$_3$, damage evolves (\eqref{eq:complementarity}$_1$ holds with a strict inequality) in  regions where the system satisfies first order optimality conditions~\eqref{eq:complementarity}$_2$ with an equality.
        At first order, this is handled by projecting the energy gradient onto the subspace orthogonal to the set of active constraints, ensuring the solver proceeds in directions that respect irreversibility.
        This projection modifies the search space as the set of active constraints evolves at each load, requiring the algorithm to re-evaluate the target function, the constraints, and the constrained subspace at each variation of damage.
        This guarantees that the damage field respects the irreversibility condition throughout the evolution.
        This robust and accurate enforcement of the first-order criticality, essential for the fidelity of phase-field models in fracture mechanics, however introduces a significant computational overhead.
        Indeed, every time a constraint activates or deactivates (e.g., when a node reaches or leaves the energy optimality threshold), the constraint matrix is reorthogonalized.
        This operation has a computational cost of $O(n+\Delta k)k$, where $n$ is the total number of degrees of freedom, $k$ is the number of active constraints, and $\Delta k$ represents the incremental changes in the active set.
        Furthermore, each evaluation of the target energy functional incurs an additional computational cost $O(n)$.
        By our choice of kinematic boundary conditions
        and the one-dimensional setting the damage criterion is attained as soon as the load is nonzero, hence this re-evaluation of constraints is performed for all nodes.
    }

To solve the second-order cone-constrained inequality~\eqref{eq:variational_stability}, {both our solvers} employ a numerical method based on the orthogonal decomposition of the Hilbert space  $X_0$  according to two mutually polar cones  $K^+_0$  and  $K^*$, cf. {\cite{Moreau1962-fz,Pinto_da_Costa2010-qv}}. Given an element  $z$  in  $X_0$, there exists a unique decomposition into two orthogonal components,  $x \in K^+_0$  and  $y \in K^*$ , where  $x$  and  $y$  are the closest elements to  $z$ in  $K^+_0$  and  $K^*$, respectively.

This decomposition allows us to project the second order problem into the cone and ensure that the eigen-solution satisfies the constraints imposed by irreversibility. This  is particularly useful in mechanics and physics when dealing with unilateral constraints or problems where the solution space is naturally bounded by physical considerations (e.g., non-negative stress, plastic deformations, etc.)

We implement a simple iterative Scaling-and-Projection algorithm~\cite{Pinto_da_Costa2010-qv} which depends upon one numerical parameter, a scaling factor $\eta>0$, {and an initial guess}. Given a convex cone  $K$ and an initial guess $z_0$ (not necessarily in the cone), the algorithm operates by first projecting the vector into the cone, $x^{(k=0)}= \operatorname{P_K}(z_0)$, then { iteratively computing the eigenvalue} estimate using the Rayleigh quotient
\begin{equation}
    \lambda^{(k)} = \frac{{x^{(k)}}^T H x^{(k)}}{||x^{(k)}||},
\end{equation}
where $H$ is the {(projected)} Hessian operator. Then, we compute the residual vector $y^{(k)} = H x^{(k)} - \lambda^{(k)} x^{(k)}$ and obtain the next iterate $x^{(k+1)} = v^{(k)}/||v^{(k)}||$ where
$v^{(k)} =\operatorname{P_K} (x^{(k)} + \eta y^{(k)})$.
The algorithm is repeated until convergence is achieved {on $x^{(k)}$}. Note that, in the cone-constrained case, the residual vector $y^{(k)}$ need not be zero at convergence.

    {We compare spectral information from the two solver setups (see Figure~\ref{fig:shouldnt}), namely:
        i) Quasi-Newton (with approximate Hessian, L-BFGS/CG): shown with large orange and blue markers,
        ii) Newton-based solver (with exact Hessian, PETSc/SNES): shown with small red and black circles.
        In both cases, we display:
        the bifurcation spectrum (blue markers, black circles), and the cone-constrained stability spectrum (orange markers, red circles).
        At small loads both solvers follow the homogeneous path, consistent with theoretical predictions and the uniqueness of the solution below the bifurcation threshold, despite the homogeneous branch being stable.
        As the load increases, however, a slight oscillation appears in the damage field before the theoretical bifurcation load $\bar{\epsilon}_b$, for the Quasi-Newton solver.
        This is  clearly appreciated evaluating the $L^2$-norm of the damage gradient, marking the onset of spatial localisation, see  the figure inset. The damage profile corresponding to the theoretical bifurcation
        already deviates slightly from homogeneity, at a load approximately 20\% smaller than the bifurcation point.
        These early oscillations are reminiscent of the $n=3$ bifurcation mode computed analytically (cf. Fig.\ref{fig:kick} ($H'\mapsto A$)). The bottom panel of Figure~\ref{fig:shouldnt} presents the spectral diagram for the irreversible problem, displaying the smallest eigenvalues of the bifurcation (ball) and stability (cone-constrained) problems.
        This {early} transition, marked by oscillations is a spurious bifurcation driven by the numerical approximation rather than a loss of physical stability.
        In contrast, the Newton-based solver correctly tracks the stability of the homogeneous solution across the bifurcation point, showing strictly positive, albeit discontinuous, cone eigenvalues throughout.
        Corresponding trajectories are shown in the equilibrium map in Figure~\ref{fig:branches-stiff}.
        This behaviour highlights the sensitivity of the system. While irreversibility implemented at first order by the active-set method guarantees damage monotonicity, the Hessian approximation performed by the Quasi-Newton method disrupts the stability information, {showing how} approximate quasi-Newton methods may drift away from physically meaningful trajectories.
    }

\begin{figure}[htbp]
    \centering
    \begin{overpic}[
            trim={0 0 1cm 2.3cm},clip,
            width=.95\textwidth
        ]{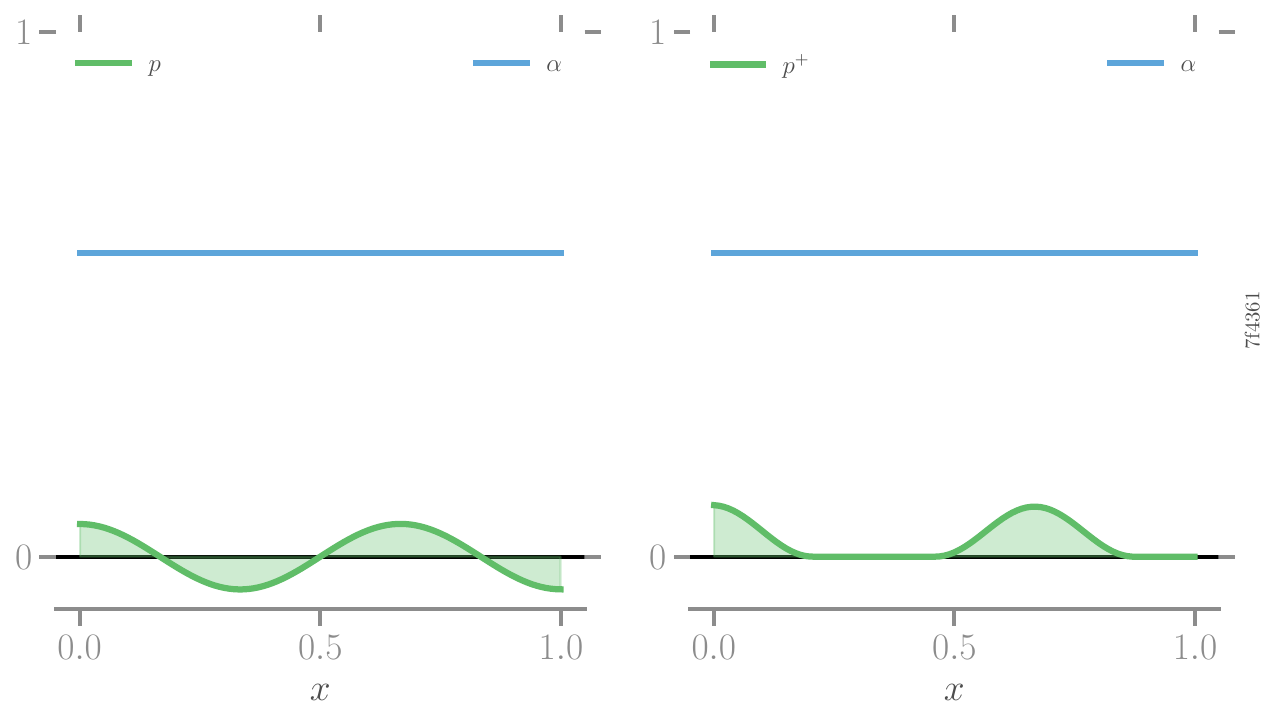}
        \put(25,40){\large $\alpha_h$}
        \put(13,17){\large $\beta \in H^1_0(0,1)$}
        \put(75,40){\large $\alpha_h$}
        \put(70,17){\large $\beta^+ \in K_0^+$}
    \end{overpic}
    \caption{{Damage field profiles (blue) and inf-eigenvectors (green) at bifurcation load $\bar{\epsilon}_b$ for the bifurcation problem (left) and the constrained stability problem (right). At bifurcation, damage $\alpha_h$ is homogeneous. The bifurcation field $\beta$ is an eigenmode on the branch $n = 4$ (cf.~Fig.~\ref{fig:branches-stiff}-right). The stability problem's inf-eigenvector ($\beta^+$ on the right) is associated to an eigenvalue $\lambda = 2 \cdot 10^{-2}>0$.} {These solution profiles respectively correspond to points $\mathbf{A, B}$ in Fig.~\ref{fig:shouldnt}(bottom)}}
    \label{fig:irreversibility-profiles}
\end{figure}

\begin{figure}[htbp]
    \centering
    \includegraphics*[width=.8\textwidth]{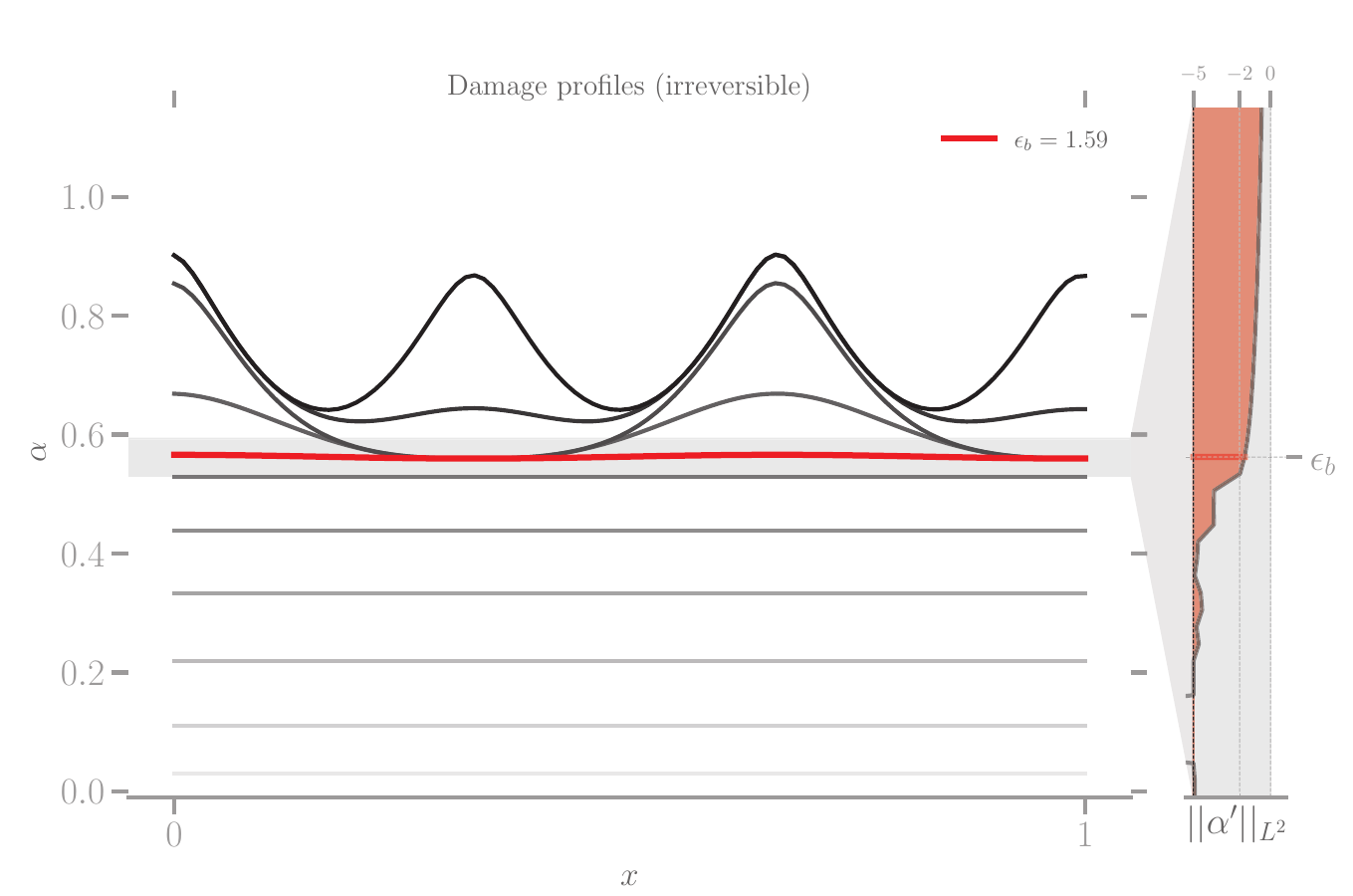}
    \includegraphics*[width=.8\textwidth]{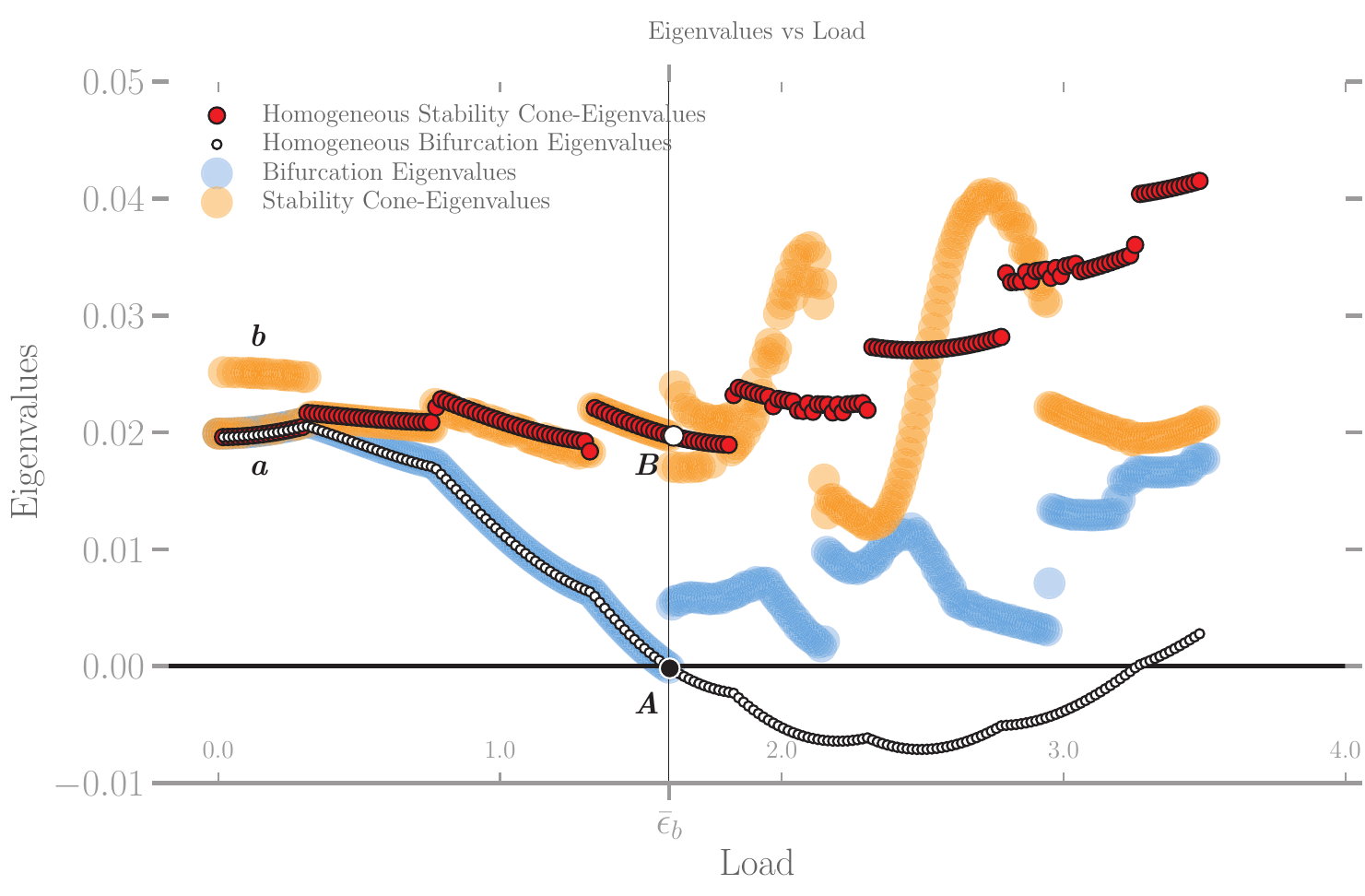}

    \caption{
        {
                \textbf{Top:}
                Damage profiles from a Quasi-Newton solver with irreversibility. At first, the solution is homogeneous (bottom, light gray). As the load increases, small perturbations (oscillations) appear before the theoretically predicted bifurcation point $\bar{\epsilon}_b$, highlighted in red. This indicates a spurious numerical bifurcation due to solver sensitivity.
                \textbf{Inset (Top Right):} As a diagnostic, the $L^2$ norm of the damage gradient highlights the emergence of spatial non-homogeneity (oscillations). The red line corresponds to the theoretical bifurcation load, tick marks indicate value of the norm (base-10 logarithmic scale).
                \textbf{Bottom:}
                We compare the smallest eigenvalues computed from two numerical approaches: a Quasi-Newton solver (visualised using large blue and orange markers) and a full Newton solver (visualised using small circles, black/white and red markers). For each method, we compute two types of spectra: the bifurcation spectrum (light blue and black/white markers), which measures the loss of uniqueness of equilibrium solutions, and the stability spectrum (orange and red markers), which reflects the system's resistance to perturbations under the irreversibility constraint.
                The two solvers agree up to the critical bifurcation load, and diverge beyond.
                The full Newton solver with exact Hessian properly captures the stability of the homogeneous solution, showing its stability even after bifurcation (indicating the solution remains observable, though non-unique).
                This discrepancy reveals that Quasi-Newton methods, due to their partial Hessian information, can trigger premature (unphysical) transitions. In contrast, Newton-type methods with full Hessian accurately preserve stability information.
                Spectral Detail: the difference between cone (stability) and ball (bifurcation) eigenvalue problems also reflects in the structure of their spectra (piecewise continuous vs. smooth). Their solutions are sampled at points A and B, corresponding to the zero (singular) bifurcation eigenmode (A, cf. Fig~\ref{fig:irreversibility-profiles}-left) and to a stable mode not causing transition (B, cf. Fig~\ref{fig:irreversibility-profiles}-right).}
    }
    \label{fig:shouldnt}
\end{figure}

    {The full Newton PETSc solver, leveraging full Hessian information, maintains fidelity to the underlying variational structure and avoids spurious instabilities.
        There is qualitative difference in the spectral regularity: the bifurcation inf-spectrum (infimum of eigenvalue across time) is continuous and piecewise smooth, while the stability inf-spectrum (cone-constrained infimum eigenvalue) is globally discontinuous.
        Our numerical experience is that the cone-constrained eigenvalue is sensitive to the initial guess.
        At low loads (the spectral branches labelled  $a$ and $b$ in Figure~\ref{fig:shouldnt}{–(bottom)}), when the initial guess lies outside the admissible cone, the projection performed to bootstrap the second order solver returns a trivial zero mode, resulting in an overestimation of the smallest eigenvalue.
        At higher loads, the bifurcation mode providing the initial guess lies closer to the admissible set, and the spectral estimate becomes more accurate.
        We sample eigenmodes at the bifurcation load $\bar{\epsilon}_b$ (marked \textbf{A} and \textbf{B} in the bottom panel of Figure~\ref{fig:shouldnt}), associated to the homogeneous state $H$ in the energy diagram in Fig.~\ref{fig:branches-stiff}.
        Their corresponding profiles are displayed in Figure~\ref{fig:irreversibility-profiles}-(left) and -right, respectively.
        Altogether, this  underscores that not all observed bifurcations in numerical simulations are physical.
        In constrained systems, numerical artefacts arising from the choice of solver or approximation strategy can spuriously trigger localisation and corrupt the stability information encoded in the (singular) energy curvatures.
        The stability and bifurcation spectra are essential tools to distinguish physical transitions from purely numerical effects.}

Finally, note that the bifurcation spectrum is singular at $\bar \epsilon_t=0$ because the damage criterion is attained as soon as the load is non-zero, and the space of admissible rate perturbations changes suddenly from $H^1_0(0,1) \times \emptyset$ at $\bar \epsilon_t=0$ to the full space $X_0 = H^1_0(0, 1)\times H^1(0, 1)$ for $\bar \epsilon_t>0$ which includes all (sufficiently smooth) damage rate perturbations.
Despite the occurrence of negative eigenvalues for the bifurcation problem, the eigenvalues of the stability problem are all positive, which is a sufficient condition to determine the stability (and thus, the observability) of the computed homogeneous evolution.


\section{Discussion}
\label{sec:discussion}
Capturing branch switching phenomena across stability transitions is not an automatic feature of approximate numerical methods. If they rely on approximate information about the Hessian of the energy functional, these methods do not guarantee to  detect transitions between critical equilbrium states, when stability is lost.
Indeed, this requires a careful determination of the zero-eigenmodes that render singular the exact nonlinear Hessian, which is typically not available in general purpose {numerical optimization} algorithms.
In practice, without such information, critical loads for equilibrium transitions become algorithm-dependant and are not consistent with closed-form solutions of the exact evolution problem.

Our evidence is that certain numerical methods can introduce non-physical artifacts which should be distinguished from genuine physical phenomena. Our ongoing work aims to refine numerical techniques to provide more reliable algorithms for analysing irreversible processes in variational evolutionary problems with multiple local minima and a high number of degrees of freedom.

From our numerical experience, first order solutions to strongly nonlinear, nonconvex, and singular problems, like those of interest in the applications, exhibit strong sensitivity to numerical errors, possibly leading to spurious bifurcations and artificial state transitions. On the other hand, solutions which integrate second order information are robust and their observability can be fully characterised.

More than numerical perturbations (which can always arise), the use of numerical methods relying only on (conjugate) gradients (in lieu of exact Hessians) is prone to introducing non-physical crack nucleation.
    {Our analysis shows that while quasi-Newton algorithms accelerate convergence, this comes at the cost of introducing unphysical effects. In particular, the approximations inherent in quasi-Newton updates may lead to solutions that appear physical yet are driven by numerical disorder rather than the governing equations.
        This sensitivity indicates that when irreversibility constraints are critical, a full Hessian approach is  necessary to accurately capture the stability and bifurcation behavior.
        Ultimately, practitioners must carefully balance the computational efficiency of quasi-Newton methods against the increased reliability (and potential cost) of employing the full Hessian.}
This is an important observation, which highlights the need for a thorough investigation of the stability of solutions.
If only physical factors are considered, an energetic selection mechanism is already encapsulated in the stability statement in the evolution law. As a consequence, equilibrium solutions under increasing load should be maintained as observable if stable, assuming that no nucleation should occur otherwise.

    {
        The brittle thin film model studied here offers a balance between analytical tractability and phenomenological complexity: its one-dimensional setting allows great simplification (due to  stresses being constant), while the elastic foundation introduces a nontrivial lower-order perturbation to the energy landscape. This interplay gives rise to a rich bifurcation structure that challenges standard numerical approaches and highlights the sensitivity of the solution process,  serving as a rigorous benchmark for assessing the impact of solver choices and stability formulations, particularly in the presence of irreversibility constraints.
        This setting allows us to rigorously examine solver-induced artefacts, the role of the Hessian approximation, and the relevance of second-order conditions in the irreversible context, all within a framework that remains accessible to detailed analysis.
        While we have considered a perfectly symmetric setting and homogeneous material, despite the existence of  modern sub-micron applications where components are effectively defect-free, we acknowledge that real systems have intrinsic oscillations.
        Small perturbations - whether from initial `defects' or material inhomogeneities - indeed influence the evolution.
        To this end, our cone-constrained stability framework provides a meaningful measure of robustness, as the lowest eigenvalue quantifies the system's spectral distance from instability.
        In our simple example, this distance remains approximately constant across the homogeneous evolution including near the bifurcation point, suggesting that the observed homogeneous solution is not fragile or unduly sensitive.
        This is why our framework is particularly conducive to further investigation of the potential impact of slight imperfections, through specialized perturbation analysis.
        Future work will include a dedicated study focusing on the impact of imperfections, noise, and material perturbations on the stability and bifurcation behaviour, as well as
        a more detailed exploration of evolutionary algorithms and their implementation for stability analysis.
    }

For interpreting {our findings} in real-world contexts  where accurate and robust algorithms are essential, such as the formation of craquelures in  paintings~\cite{fuster-lopez:2020-picassos, Bosco:2020aa,Bosco:2021}, damage and breakup of brittle ice structures~\cite{weiss:2017-linking, tollefson:2017-giant, Sun:2023aa, Millan:2023aa} and crack-pattern selection in metallic thin films at small-scale~\cite{Faurie2019-to}, we outline two principal strategies:

\begin{enumerate}
    \item
          Ignore the Numerical Artifact: focussing solely on first order considerations and acknowledging that the observed computed nucleations may be purely numerical and should not be considered in physical terms.
    \item
          Highlight the Numerical Artifact: emphasizing that stable solutions should be observable, despite
          the sensitivity to numerical parameters (artifacts, in the quasi-Newton approach) and the abundance of admissible solutions (in the nonconvex scenario), or
    \item
          Otherwise.
\end{enumerate}

Suggesting that state transitions in complex scenarios should be carefully interpreted, the connection between observability and stability is functional to understanding real patterns that emerge, e.g., in higher dimensions or in other physical systems.

In either case, our computations show that, unless second order analysis is performed, observed nucleations are \emph{not necessarily} indicative of physical cracks but rather of an interplay between purely  {physical} phenomena, inherent to the nature of mechanical processes, and numerical biases inherent to the computational methods employed.
This distinction is crucial for understanding the limitations and proper application of numerical techniques as predictive tools in contexts where cracks are a real concern for structures.

\paragraph{Acknowledgements} This work was in part supported by the French National Research Agency (ANR, Project No. ANR-20-CE91–0010) and the Austrian Science Fund (FWF, Project No. I 4913-N) within the framework of the project: Nanoarchitected films for unbreakable flexible electronics (NanoFilm).
\end{document}